\documentclass[acmtog]{acmart}
\acmSubmissionID{113}

\usepackage{booktabs} 

\usepackage[ruled]{algorithm2e} 

\SetAlFnt{\small}
\SetAlCapFnt{\small}
\SetAlCapNameFnt{\small}
\SetAlCapHSkip{0pt}

\acmJournal{TOG}

\setcopyright{acmcopyright}\acmJournal{TOG}
\acmYear{2020}\acmVolume{39}\acmNumber{6}\acmArticle{228}\acmMonth{12}
\acmDOI{10.1145/3414685.3417764}

\usepackage{multirow}
\usepackage{caption}
\usepackage{subfig}

\begin{document}
\title{Mononizing Binocular Videos}

\newcommand{\new}[1]{{{#1}}}
\newcommand{\review}[1]{{\color[RGB]{240 128 128}{[Review comment: #1]}}}
\newcommand{\ttw}[1]{{\color{blue}#1}}
\newcommand{\phil}[1]{{\color[rgb]{0.3,0.8,0.3}#1}}
\newcommand{\mhx}[1]{{\color[rgb]{0.7,0.1,0.7}#1}}
\newcommand{\wbh}[1]{{\color[rgb]{0.6,0.1,0.5}#1}}

\author{Wenbo Hu}

\author{Menghan Xia}

\author{Chi-Wing Fu}

\author{Tien-Tsin Wong}
\affiliation{%
  \institution{The Chinese University of Hong Kong}
  \city{Hong Kong}}
\email{[wbhu,mhxia,cwfu,ttwong]@cse.cuhk.edu.hk}

\begin{abstract}
This paper presents the idea of {\em mono-nizing\/} binocular videos and a framework to effectively realize it. 
Mono-nize means we purposely convert a binocular video into a regular monocular video with the stereo information implicitly encoded in a visual but nearly-imperceptible form. 
Hence, we can impartially distribute and show the mononized video as an ordinary monocular video. 
Unlike ordinary monocular videos, we can restore from it the original binocular video and show it on a stereoscopic display.
To start, we formulate an encoding-and-decoding framework with the pyramidal deformable fusion module to exploit long-range correspondences between the left and right views, a quantization layer to suppress the restoring artifacts, and 
the compression noise simulation module to resist the compression noise introduced by modern video codecs.
Our framework is self-supervised, as we articulate our objective function with loss terms defined on the input: 
a monocular term for creating the mononized video, an invertibility term for restoring the original video, and a temporal term for frame-to-frame coherence.
Further, we conducted extensive experiments to evaluate our generated mononized videos and restored binocular videos for diverse types of images and 3D movies.
Quantitative results on both standard metrics and user perception studies show the effectiveness of our method.
\end{abstract}

%
%
%

%
%


\begin{teaserfigure}
	\centering
	\includegraphics[width=0.99\linewidth]{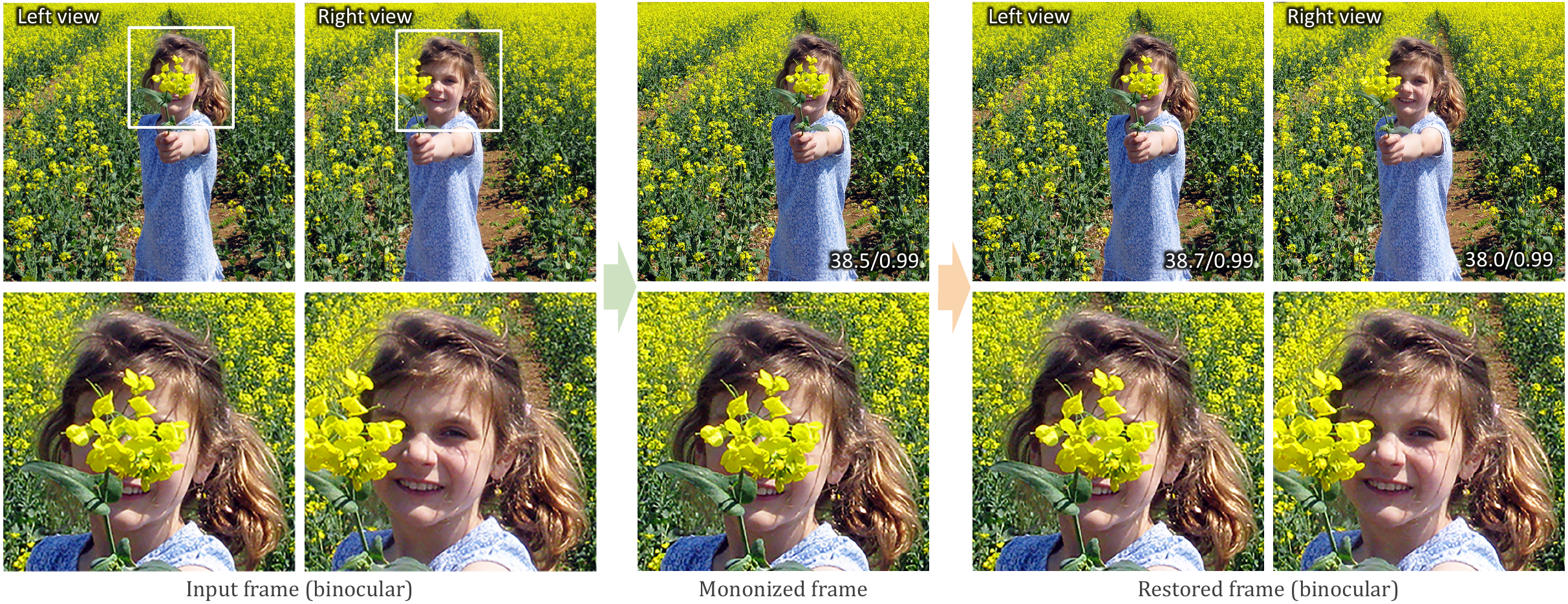}
	\vspace*{-2.0mm}
	\caption{Our method can effectively {\em mono-nize\/} a binocular video into a monocular video with the stereo information encoded in a 
	nearly-imperceptible form.
	Note, the face of the little girl in the input right view (left); we can fuse it with the left view and encode (hide) it in the mononized frame (middle).
	Though the face is not observable in the mononized frame, we can restore it back in the restored binocular frame (right).
	We show PSNR/SSIM at the bottom of each result.
}
	\label{fig:teaser}
	\vspace*{1.0mm}
\end{teaserfigure}

\maketitle


\section{Introduction}
\label{sec:introduction}
While multi-camera smartphones become popular in the market,
%
single-view platforms remain dominant. 
To migrate from single- to multi-view, re-design and re-implementation of existing software and hardware platforms are usually unavoidable, and may obsolete the existing single-view platforms.
Hence, a backward-compatible solution is crucial, as demonstrated in the successful migration from black\&white to color TV broadcasting during the 50's to 60's.

%
%
%
%
%

%
%
%
%

\begin{figure}[!t] 
	\centering
	\includegraphics[width=0.99\linewidth]{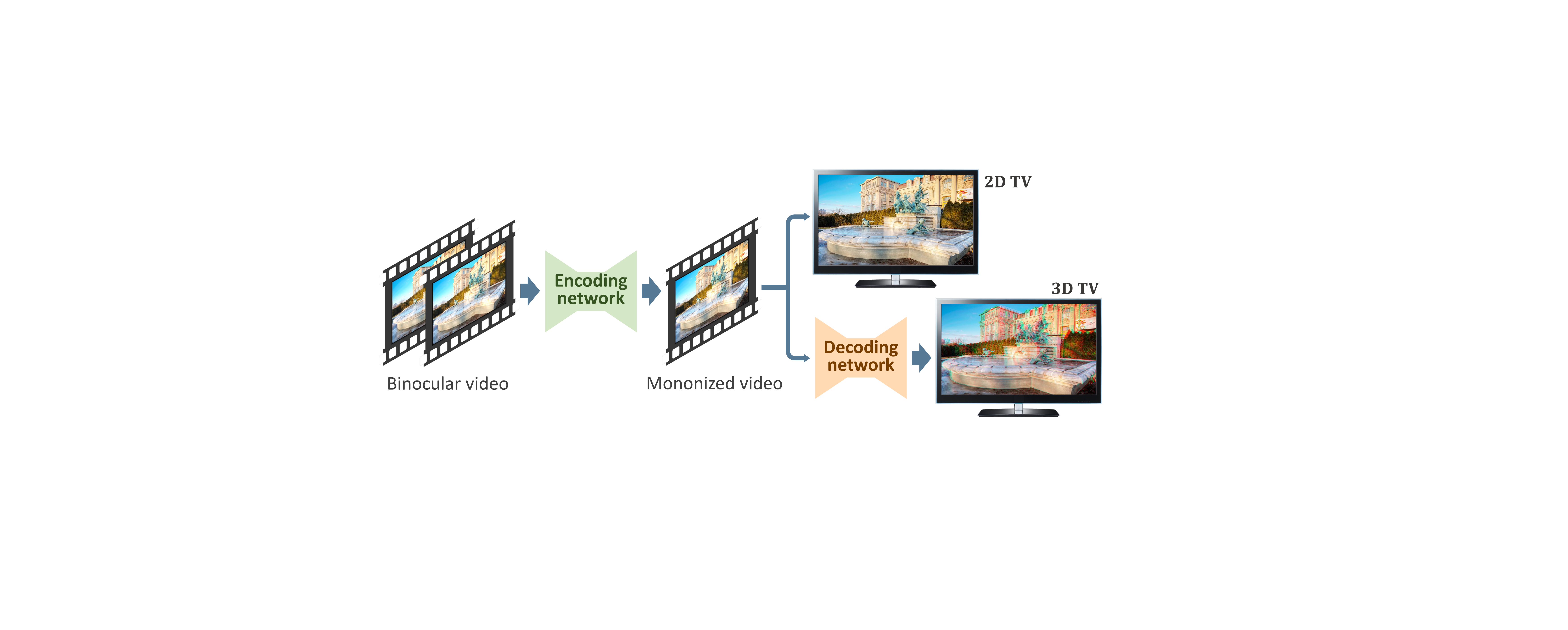}
	\vspace{-1.5mm}
	\caption{
	Given a binocular video, our framework produces a mononized video that is visually no different from an ordinary monocular video, and can be distributed and displayed on conventional monocular-video platforms. 
	If a 3D display is available, we can restore the original binocular video from the mononized video and provide stereo viewing on the video.}
	\label{fig:key_idea}
\end{figure}

In this paper, we present a fully backward-compatible solution that is independent of the video coding standard and requires zero additional upgrade/installation on monocular TVs to cope with the stereoscopic data.
To achieve the goal, we propose a brand new approach to {\em ``mono-nize''\/} (convert) conventional binocular (stereoscopic) images/videos to monocular ones. 
The generated monocular images/videos, which we call {\em mononized images/videos\/}, look visually the same as one of the two views (say, the left view, without loss of generality) in the given binocular images/videos, and can be treated (stored, distributed, and displayed) as ordinary monocular images/videos. 
The only difference is that we can restore their binocular counterparts from them, whenever necessary.
The mononization and restoration enable a fully backward-compatible solution for the migration from single- to multi-view (Figure~\ref{fig:key_idea}).

A possible solution 
is to just drop one of the two views, then use methods such as~\cite{Shih3DP20,NiklausMYL19,Leimkuehler2018TVCG,calagari2017data,XieG16} to infer or estimate the other view from the remaining one.
%
However, such approach cannot accurately predict the dropped view, due to the loss of occlusion and depth information.
For instance, the face of the little girl shown in Figure~\ref{fig:teaser} is mostly occluded in the left view; if we drop the right view, it will be hard to recover her face in the right view solely from the left one.
%
Rather, we aim to implicitly encode the other view in a {\em visual but nearly-imperceptible form\/} in the mononized frame, such that we can restore from it a high-quality binocular frame (Figure~\ref{fig:teaser}).
The key problem is how to achieve such an encoding.

Instead of handcrafting the encoding process, we propose to embed the encoding system via a convolutional neural network (CNN)~\cite{Goodfellow2016}.
%
Our framework consists of (i) an {\em encoding neural network\/} to convert the input binocular video into a mononized video and (ii) a {\em decoding neural network\/} to restore the binocular video (Figure~\ref{fig:key_idea}).
This formulation enables self-supervised learning and bypasses the need of preparing manually-labeled training data, which is a common burden to many deep learning methods.

Unfortunately, it is hard for conventional CNNs to exploit correspondences between the left and right views, since the large disparities between views may exceed the spatial transformation capability of conventional CNNs~\cite{jaderberg2015spatial}.
To overcome this, we present the pyramidal deformable fusion (PDF) module to explore long-range correspondences between the left and right views.
Also, we adopt a quantization layer to suppress the 
artifacts caused by quantization errors,
and formulate the compression noise simulation (CNS) module to resist the compression perturbation that could be introduced by 
the 
video codecs.
%
Further, we design an objective function with three loss terms to train the networks:
{\em monocular loss\/} to ensure the mononized video looks like the left view of the input video,
{\em invertibility loss\/} to ensure the restored binocular video looks like the original,
and {\em temporal loss\/} to ensure the temporal coherence 
in both the mononized videos and restored binocular videos.

To evaluate our method, we employed a collection of binocular images and 3D movies of various scene categories, and conducted extensive experiments, including
%
%
a qualitative evaluation on the visual quality of our results, i.e., the mononized and restored binocular images/videos (Section~\ref{subsec:qualitative_evaluation}),
quantitative comparisons with methods~\cite{baluja2017hiding,XiaL18,NiklausMYL19} related to our application (Section~\ref{subsec:compare_sota}),
a quantitative evaluation of our results on frame quality and temporal coherence (Section~\ref{subsec:quantitative_evaluation}),
a quantitative evaluation on the compatibility of our method with common video codecs (Section~\ref{subsec:video_codecs}); and
a user study 
to evaluate the perceptual performance of our method (Section~\ref{subsec:user_study}).
All evaluated metrics confirm the effectiveness of our method to produce high-quality mononized and restored binocular images/videos.
Also, the user study shows that our generated mononized videos and restored binocular videos look no different from their original counterparts.
Further, the experimental results show that our mononized videos are friendly with common video codecs.
In fact, encoding our mononized videos with standard codecs outperforms existing side-by-side and multi-view encoding methods at low bit-rate, while it achieves comparable compression performance at high bit-rate.
This makes our approach a favorable alternative when storage is a concern.


Our contributions are summarized below.
\begin{itemize}

\vspace*{-1.5mm}
\item We offer a 
backward-compatible solution for distributing and storing binocular videos as monocular ones, that are fully compatible to common monocular platforms.


\item We formulate this conversion-and-restoration problem as an encoding-and-decoding process embedded in deep neural networks that are trained in a self-supervised manner.

\item We propose the pyramidal deformable fusion module to exploit the long-range correspondences between the left and right views, 
design a compression noise simulation module and adopt a quantization layer 
to resist noise in real-world video compression.

\item We offer a favorable compression alternative, as our method outperforms existing solution at low bit-rate, and achieves a comparable performance at high bit-rate.

\end{itemize}

\section{Related Work}
\label{sec:relatedWorks}
\subsection{Stereo Image/video Synthesis}
\label{subsec:stereo_synthesis}
Novel view synthesis from a single image, e.g.,~\cite{Leimkuehler2018TVCG,NiklausMYL19}, usually starts by estimating a depth map, then performing depth-based image rendering.
%
Deep neural networks have shown remarkable improvements on single-image depth estimation~\cite{EigenP14,atapour2018real,luo2018single,facil2019cam,li2019learning}.
However, it remains 
very challenging to obtain accurate depth maps for general scenes.
Several works~\cite{XieG16,cun2018depth,liu2018geometry} propose to integrate depth estimation and view synthesis into an end-to-end deep neural network.
Recent works also propose to explicitly inpaint the occluded regions in image~\cite{Shih3DP20} or point cloud domain~\cite{NiklausMYL19}.
However, ensuring plausible results in the inpainted occluded regions is still very hard.
%
%
%
Instead of estimating the novel views, this work solves a very different problem of {\em restoring\/} the stereo contents in the mononized video, in which the stereo information is implicitly encoded.

Besides, some works focus on retargeting stereo or multi-view images with preferred properties, e.g., stereo magnification~\cite{zhou2018stereo}, novel view synthesis from multiple images~\cite{FlynnN16,xu2019deep,Lombardi2019,srinivasan2019pushing,kellnhofer20173dtv,mildenhall2020nerf}, and disparity manipulation~\cite{LangH10,DidykR11,KellnhoferD16,didyk2012luminance}.
Recently, some works~\cite{ScherL13,FukiageK17} point out that when a conventional stereoscopic display is viewed without stereo glasses, image blurs, or `ghosts,' are visible due to the fusion of the stereo image pairs, which makes the stereoscopic display backward incompatibility.
Also, they propose to synthesize ghost-free stereoscopic image pairs that can minimize the ghosts when viewed without stereo glasses and provide stereo viewing when stereo glasses is available.
Note that ghost-free stereo images/videos are still binocular in nature.
In contrast, our mononized images/videos are monocular in nature, and yet embed the stereo information in a nearly-imperceptible form.

\subsection{Reversible Image Conversion}
\label{subsec:invertible_conversion}
The reversible property has been explored in the stenography methods~\cite{baluja2017hiding,Zhu18hidden,WangGWXZL19,wengrowski2019light}, which conceal secret information (or an image) within a reversible container image, from which the secret information can be recovered.
For example, Baluja~\shortcite{baluja2017hiding} presents a method to hide an image within another one.
However, the container image and secret information/image are usually unrelated, so it is hard to generate artifacts-free results for both hiding and recovering.

On the other hand, 
the reversible property is explored in some image processing procedures.
%
For example, \cite{XiaL18} formulate a neural network to generate a reversible grayscale from a color image, where colors can be restored from the grayscale image;
\cite{LiYue19} adopt a neural network to generate down-sampled images with compactly-encoded higher-resolution details, and show that the high-resolution counterparts can be restored with a super-resolution neural network.
Our work belongs to the category of reversible networks.
%
%
Among them,~\cite{XiaL18} is a generic one.
However, we cannot directly apply it to realize our application.
Its network is designed for still images.
Also, it cannot effectively hide a view (right view) in the other one (left view), since the large disparities between views may exceed the spatial transformation capability of conventional CNNs.
Moreover, we need to handle binocular videos and account for the temporal coherence.
Please refer to Section~\ref{subsec:compare_sota} for comparison with other methods.
\subsection{Binocular Video Encoding}
\label{sec:codec}
To store and distribute a binocular video using existing video encoding standards, one can simply concatenate each pair of left and right frames top-down or side-by-side as a single image~\cite{vetro2010frame}, and regard the combined frames as a regular monocular video for encoding using existing video codecs.
Doing so not only ignores the binocular coherence in the encoding, but also wastes the computation, if we simply want to playback the video on conventional monocular displays.
It is because we first have to decode the whole side-by-side video before obtaining the individual frames and dropping the other half of the frames; we cannot skip the decoding of the other half, due to the nature of video encoding.
Having said that, we need a special piece of software/hardware to handle the process, which is incompatible with regular monocular displays.

Another stream of approaches is to design multi-view extensions for existing monocular video codecs, 
e.g., the MVC extension\footnote{\url{https://en.wikipedia.org/wiki/Multiview_Video_Coding}} for H.264/MPEG4-AVC uses 2D plus Delta to encode binocular videos
%
and the MV-HEVC\footnote{\url{https://hevc.hhi.fraunhofer.de/mvhevc}} extension for H.265/HEVC supports the encoding of multiple views with inter-layer prediction.
More multi-view extensions for H.264/MPEG4-AVC and for H.265/HEVC can be found in~\cite{vetro2011overview} and~\cite{tech2016overview}, respectively.
However, multi-view extensions work for specific codecs: each time a new codec comes out, the multi-view extension has to be re-designed and re-developed.
Recently,~\cite{mallik2016hevc} and~\cite{lai2017content} present new multi-view extensions for H.265/HEVC that use 4D wavelets and frame interleaving to enhance the coding efficiency.
%
Unfortunately, they are 
incompatible with diverse existing monocular displays.
Different from the above methods, our framework does not rely on any specific video codec.
We can readily restore the binocular video from the mononized one via a decoding network on top of the video codec, so our framework is fully compatible with the existing codecs, as demonstrated in Section~\ref{subsec:video_codecs}.

\newcommand{\IL}{$\mathbf{I}_L$}
\newcommand{\IR}{$\mathbf{I}_R$}
\newcommand{\OL}{$\mathbf{O}_L$}
\newcommand{\OLt}{$\mathbf{O}_L^{\ t}$}
\newcommand{\OLtminusone}{$\mathbf{O}_L^{\ t-1}$}
\newcommand{\OR}{$\mathbf{O}_R$}
\newcommand{\ORt}{$\mathbf{O}_R^{\ t}$}
\newcommand{\ORtminusone}{$\mathbf{O}_R^{\ t-1}$}
\newcommand{\OM}{$\mathbf{O}_M$}
\newcommand{\OMt}{$\mathbf{O}_M^{\ t}$}
\newcommand{\OMtminusone}{$\mathbf{O}_M^{\ t-1}$}
\newcommand{\PL}{$\mathbf{P}_{L}$}
\newcommand{\PLt}{$\mathbf{P}_{L}^{\ t}$}
\newcommand{\PR}{$\mathbf{P}_{R}$}
\newcommand{\PRt}{$\mathbf{P}_{R}^{\ t}$}
\newcommand{\PM}{$\mathbf{P}_{M}$}
\newcommand{\PMt}{$\mathbf{P}_{M}^{\ t}$}
\newcommand{\PMtminusone}{$\mathbf{P}_{M}^{\ t-1}$}
\newcommand{\PMtilde}{$\widetilde{\mathbf{P}_{M}}$}
\newcommand{\PMtildet}{$\widetilde{\mathbf{P}_{M}^{\ t}}$}
\newcommand{\PLtilde}{$\widetilde{\mathbf{P}_{L}}$}
\newcommand{\PLtildet}{$\widetilde{\mathbf{P}_{L}^{\ t}}$}
\newcommand{\PLtildetminusone}{$\widetilde{\mathbf{P}_{L}^{\ t-1}}$}
\newcommand{\PRtilde}{$\widetilde{\mathbf{P}_{R}}$}
\newcommand{\PRtildet}{$\widetilde{\mathbf{P}_{R}^{\ t}}$}
\newcommand{\PRtildetminusone}{$\widetilde{\mathbf{P}_{R}^{\ t-1}}$}


\begin{figure*}[!t] 
	\centering
	\includegraphics[width=0.999\linewidth]{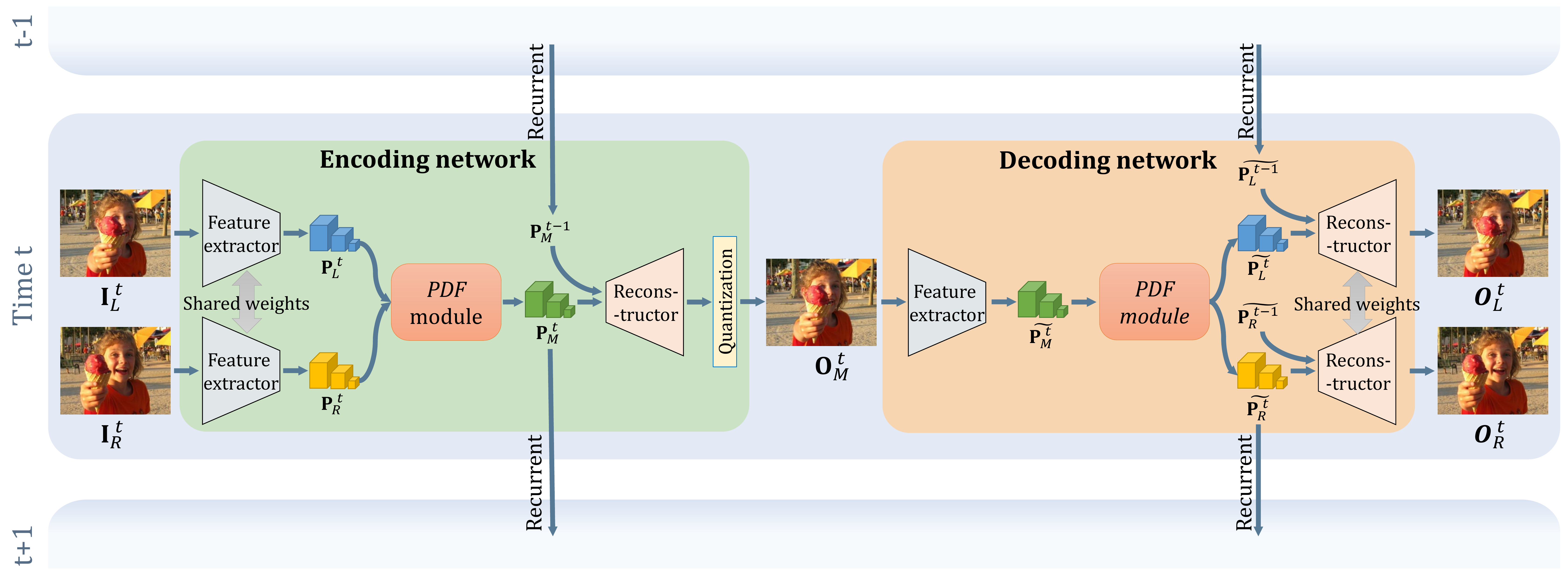}
	\vspace*{-1mm}   
	\caption{Overview of our framework for producing mononized videos.
Our framework has two parts:
(i) an encoding network that produces mononized frame \OMt{} at time $t$ from input binocular image pair \{$\mathbf{I}_L^{\ t}$, $\mathbf{I}_R^{\ t}$\} and pyramidal mononized feature \PMtminusone{} from the previous time frame; and
(ii) an decoding network that restores a binocular image pair \{$\mathbf{O}_L^{\ t}$, $\mathbf{O}_R^{\ t}$\} from $\mathbf{O}_M^{\ t}$ and binocular pyramid feature pair \{\PLtildetminusone, \PRtildetminusone\} from the previous time frame.
Note that \PLt, \PRt{}, and \PMt{} denote the pyramidal left, right, and mononized features, respectively;
%
and the pyramidal deformable fusion (PDF) module is proposed to exploit the long-range correspondences between the left and right views to improve the encoding efficiency.
}
	\label{fig:overview}
\end{figure*}

\section{Overview}
\label{sec:overview}

\paragraph{Image Mononization.}
Before we present the full framework (Figure~\ref{fig:overview}) for producing \textit{mononized videos}, we first introduce our framework for producing \textit{mononized images} to better state the insight in our approach and to give the notations.

Overall, our framework has two parts: {\em an encoding neural network\/} $E$ and {\em a decoding neural network\/} $D$, as shown in the middle ``Time $t$'' block of Figure~\ref{fig:overview} without the recurrent connections.
Given a stereo image pair \{\IL, \IR\} as input, the encoding network generates mononized image \OM{} that looks like \IL{} (without loss of generality),  and at the same time, embeds \IR{} as nearly-imperceptible information in \OM. 
Inside the decoding network, we first use a pair of feature extractors with shared weights to simultaneously extract \textit{pyramidal left feature} \PL{} and \textit{pyramidal right feature} \PR,
which are then fused together by the \emph{pyramidal deformable fusion (PDF)} module to produce the \textit{pyramidal mononized feature} \PM.
Finally, we feed \PM{} into the reconstructor to produce the mononized image \OM.

On the other hand, the decoding network restores a stereo image pair \{\OL, \OR\} from \OM, such that \{\OL, \OR\} look like \{\IL, \IR\}, respectively.
Inside the decoding network, we first extract \textit{pyramidal mononized feature} \PMtilde,
and transform it to simultaneously produce \textit{pyramidal left feature} \PLtilde{} and \textit{pyramidal right feature} \PRtilde{} by another \emph{PDF} module.
Finally, we feed \PLtilde{} and \PRtilde{} into a pair of reconstructors with shared weights to generate \OL{} and \OR, respectively.
Note, we drop superscript $t$ in the notations, since we now discuss the framework for image mononization.
Later, we will put superscript $t$ back to the notations when we discuss video mononization.
Mathematically, $E$ and $D$ are defined, respectively, as
\begin{eqnarray}
\label{eq:O_M}
\mathbf{O}_M
& = &
E( \ \{ \ \mathbf{I}_L \ , \ \mathbf{I}_R \ \} \ )
\\
\text{and} \ \
\label{eq:O_LR}
\{ \ \mathbf{O}_L \ , \ \mathbf{O}_R \ \}
& = &
D( \ \mathbf{O}_M \ )
\ \ = \ \
D( \ E( \ \{ \ \mathbf{I}_L \ , \ \mathbf{I}_R \ \} \ ) \ ) \ .
\end{eqnarray}

To train $E$ and $D$, we define the {\em monocular loss\/} $\mathcal{L}_M$ to ensure \OM{} looks like \IL{} and the {\em invertibility loss\/} $\mathcal{L}_I$ to ensure \{\OL, \OR\} look like \{\IL, \IR\}, respectively.
Altogether, we jointly train the two networks to produce a pair of compatible encoder and decoder.

\paragraph{Video Mononization}
Independently processing video frames at each time $t$ may lead to temporal incoherence.
So, we recurrently feed \PMtminusone{} and \{\PLtildetminusone, \PRtildetminusone\} from previous time $t-1$ into the networks, and formulate the {\em temporal loss\/} $\mathcal{L}_T$
to drive the network to produce \OMt, \OLt, and \ORt{} that are temporally coherent with \OMtminusone, \OLtminusone, and \ORtminusone, respectively.
Lastly, we put together the three terms to formulate the overall loss function $\mathcal{L}$, and train the whole framework end-to-end in a self-supervised manner:
\begin{equation}
\label{equation:lossForVideo} 
\mathcal{L} = \mathcal{L}_M + \lambda_1\mathcal{L}_{I} + \lambda_2\mathcal{L}_T \ ,
\end{equation}
where $\lambda_1$ and $\lambda_2$ are weights.
The details on the network architecture, the design of the loss terms, and the training scheme are presented in Sections~\ref{sec:network},~\ref{sec:loss}, and~\ref{sec:training}, respectively.

\section{Network Architecture}
\label{sec:network}
\subsection{Network Backbone}
\label{sec:backbone}


\begin{figure}[!t] 
	\centering
	\includegraphics[width=0.99\linewidth]{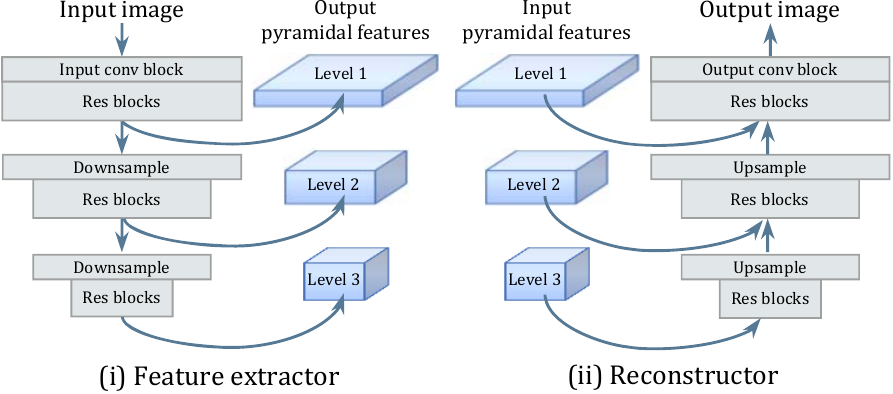}
	\vspace*{-2mm}
	\caption{
		The feature extractor (i) hierarchically extracts features in three levels, whereas 
		the reconstructor (ii) is symmetric with the feature extractor and trainable skip connections are adopted.
		%
}
 	\label{fig:network}
 	\vspace*{-1mm}
\end{figure} 

The feature extractors in both the encoding and decoding networks have the same architecture (Figure~\ref{fig:network} (i)), which is a variant of ResNet~\cite{he2016deep}.
%
We remove the batch normalization~\cite{IoffeS15} from the original residual blocks~\cite{he2016deep} as done in~\cite{NahK17}, since we found it performs better empirically.
Also, to better abstract the features from low to high levels, we adopt a convolution with a stride of two to realize the downsampling instead of using max or average pooling.
%
We hierarchically extract information from the input image in three levels to form pyramidal features.
Intuitively, the feature extractors (left part in Figure~\ref{fig:overview}) are mapping functions that embed the left and right images to feature space, so we share the weights of the two feature extractors in the encoding network to maintain their mapping uniformity.
%
%
%

The reconstructors in the encoding and decoding networks also have the same architecture (Figure~\ref{fig:network} (ii)), which is symmetric with the feature extractor architecture.
Specifically, the ``upsample'' block is achieved by a bilinear interpolation followed by a convolution.
Except for the features in the third level, features in all the other levels are fed into the reconstructor using skip connection, then linearly combined with the upsampled features from a higher level via some trainable coefficients.
Similarly, we share the weights of the reconstructors for the left and right views in the decoding network (right part in Figure~\ref{fig:overview}).
When extending this encoding-and-decoding framework to mononize binocular videos, we further feed the corresponding pyramid features from the previous time frame (\PMtminusone, \PLtildetminusone, and \PRtildetminusone) into the corresponding reconstructors to form a recurrent neural network (Figure~\ref{fig:overview}).
Please refer to Supplemental material Section 1 for the detailed network architecture.


\subsection{Pyramidal Deformable Fusion (PDF) Module}
\label{sec:pdf}

\begin{figure}[!t] 
	\centering
	\includegraphics[width=0.99\linewidth]{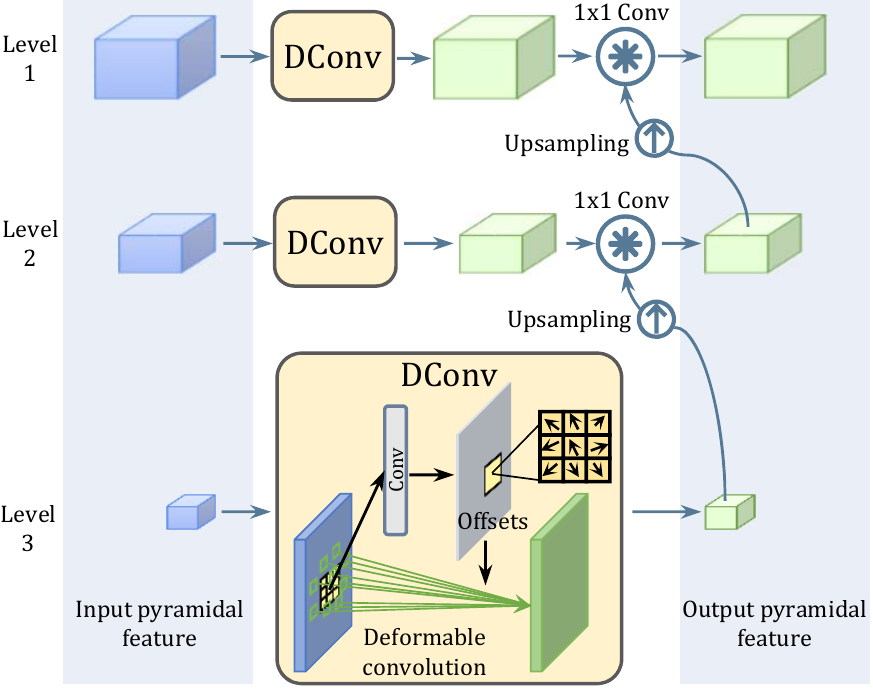}
	\vspace*{-1.5mm}
	\caption{The Pyramidal Deformable Fusion (PDF) module. Given the input pyramidal feature, the PDF module exploits the long-range correspondences and aggregates context information to generate the output pyramidal feature. The upsampling is achieved by a bilinear interpolation.
	}
	\label{fig:pdf}
	\vspace*{-2mm}
\end{figure}

First of all, the mononized pyramid feature \PMt{} should contain information of both the left and right views (left part in Figure~\ref{fig:overview}), such that the mononized frame \OMt{} reconstructed from it can inherit the information of both the left and right views and the decoding network can later extract pyramidal feature \PMtildet{} and further reconstruct the left and right views (right part in Figure~\ref{fig:overview}).

Hence, we \new{should} first fuse pyramidal features \PLt{} and \PRt{} from the left and right views to form \PMt{}.
However, \PLt{} and \PRt{} may not align well due to the disparity between the left and right views.
In practice, the disparity can be as large as $\sim$300 pixels for objects that are close to the camera.
For such cases, it will be hard for the CNNs to figure out the long-range correspondences, due to the limited spatial transformation capability
in conventional CNNs~\cite{jaderberg2015spatial}.
Similar challenge also exists in several image recognition tasks, e.g., semantic segmentation and object detection, in which 
the geometry deformation
could lead to 
performance degeneration.
%
To meet this challenge,~\cite{dai2017deformable} propose a deformable convolution operator to augment the spatial sampling locations with additional offsets and learn the offsets for semantic segmentation and object detection.
Later, the deformable convolution v2~\cite{zhu2019deformable} further extends the operator to improve the performance.
Based on deformable convolution v2, we formulate the \emph{Pyramidal Deformable Fusion (PDF)} module 
to examine the transformation between left and right views and exploit the long-range correspondences between views in a hierarchical manner.

After we concatenate the left and right pyramidal features \new{along the channel dimension at each level}, our \emph{PDF} module implicitly explores the long-range correspondences among the feature channels to produce the fused pyramidal feature, as shown in Figure~\ref{fig:pdf}.
Starting from the pyramidal feature at the third level, the \emph{PDF} module first learns the offsets from the feature map, then applies the learned offsets to the deformable convolution and produces the output feature in the same level.
For the second and first levels, we first obtain the convolved features through the same procedures as in the third level, and later aggregate the result with the bilinearly-upsampled feature from a higher level via a $1\times1$ convolution to produce the final output feature in the level.
\new{Note that, the employed deformable convolutions are with an offset group number of two, so the learned offset vectors for the pyramidal left and right features can be different.
In other words, the sampled spatial locations can be different for the left and right feature maps.}
Altogether, we produce a three-level pyramidal feature from the input three-level pyramid features.

Without changing the feature dimension and size, our \emph{PDF} module can be viewed as a feature transformation module, which exploits long-range correspondences and aggregates context information.
Hence, it can also be applied to transform \PMtildet{} back to \PLtildet{} and \PRtildet \new{with the newly learned offsets}.

\subsection{Quantization Layer}
\label{sec:quantization}

\begin{figure}[!t] 
	\centering
	\includegraphics[width=0.99\linewidth]{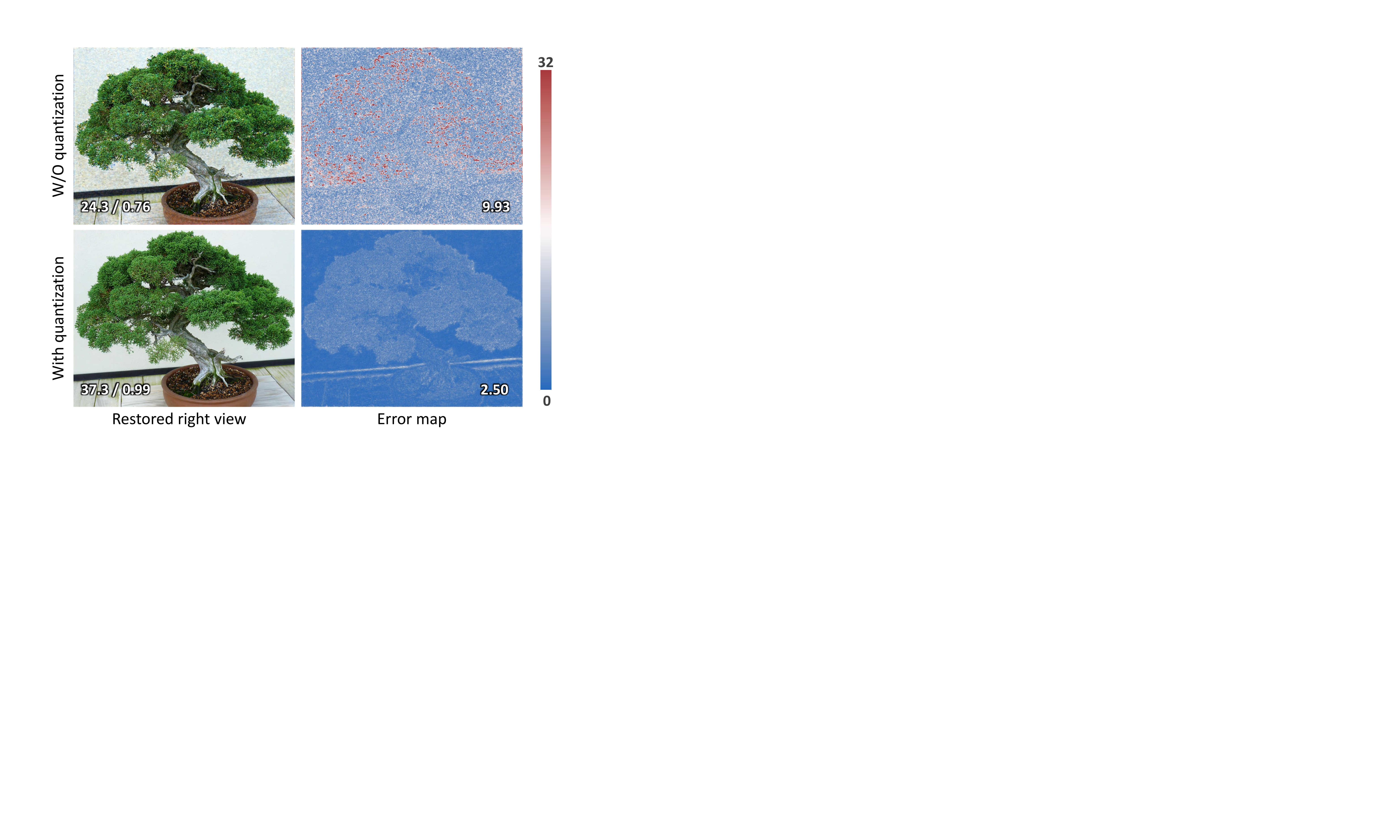}
	\vspace*{-1.5mm}
	\caption{
		Restored right views produced at the test phase using our network models trained without (top row) and with (bottom row) the quantization layer. 
		Corresponding error maps (compared with the ground truth) are shown on the right-hand side, PSNR/SSIM (left column) and mean absolute error (right column) are shown inside the restored images.
	}
	\label{fig:quantization}
	\vspace*{-2mm}
\end{figure}

The mononized frame $\mathbf{O}_M^{\ t}$ in our framework is a regular monocular image in 8-bit pixel format per RGB channel.
This means that we need to quantize the 32-bit floating point network-output values to 8-bit integers for producing $\mathbf{O}_M^{\ t}$.
Such an operation is, however, not differentiable, since it hinders the network training with gradient descent.
If we directly ignore the quantization process during the network training, the restored results could contain artifacts at the test phase (top row of Figure~\ref{fig:quantization}) due to the quantization error in the mononized view.
Inspired by the works on propagating gradients through binarization~\cite{CourbariauxH16,RastegariO17}, \new{image and network compression~\cite{agustsson2017soft}} and entropy coding~\cite{BalleLS17,choi2019variable}, we adopt a quantization layer (Figure~\ref{fig:network} (ii)), which consists of quantization function $Q(x_{ijk})$ and proxy function $\widetilde{Q}(x_{ijk})$:
\begin{equation}\label{Eq:proxy_function}
\left\{
\ \ 
\begin{aligned}
Q(x_{ijk})
& \ = &
\hspace*{-2mm}
round(x_{ijk}) \
\\
\widetilde{Q}(x_{ijk})
& \ = &
\hspace*{-1.5mm}
x_{ijk} \ ,
\\
\end{aligned}
\right.
\end{equation}
where $Q(x_{ijk})$ is used in the forward pass and the gradient of $\widetilde{Q}(x_{ijk})$ is used in the backward propagation.
By training the network with this quantization layer, we can better suppress the artifacts in the restored binocular frames (bottom row of Figure~\ref{fig:quantization}).
Also, we explored other quantization strategies, e.g. universal quantization~\cite{choi2019variable}, to build the quantization layer, and found no significant difference in the performance.
Please refer to Supplemental material Section 3 for the details.


\subsection{Compression Noise Simulation (CNS) Module}
\label{sec:cns}

\begin{figure}[!t]
	\centering
	\includegraphics[width=0.99\linewidth]{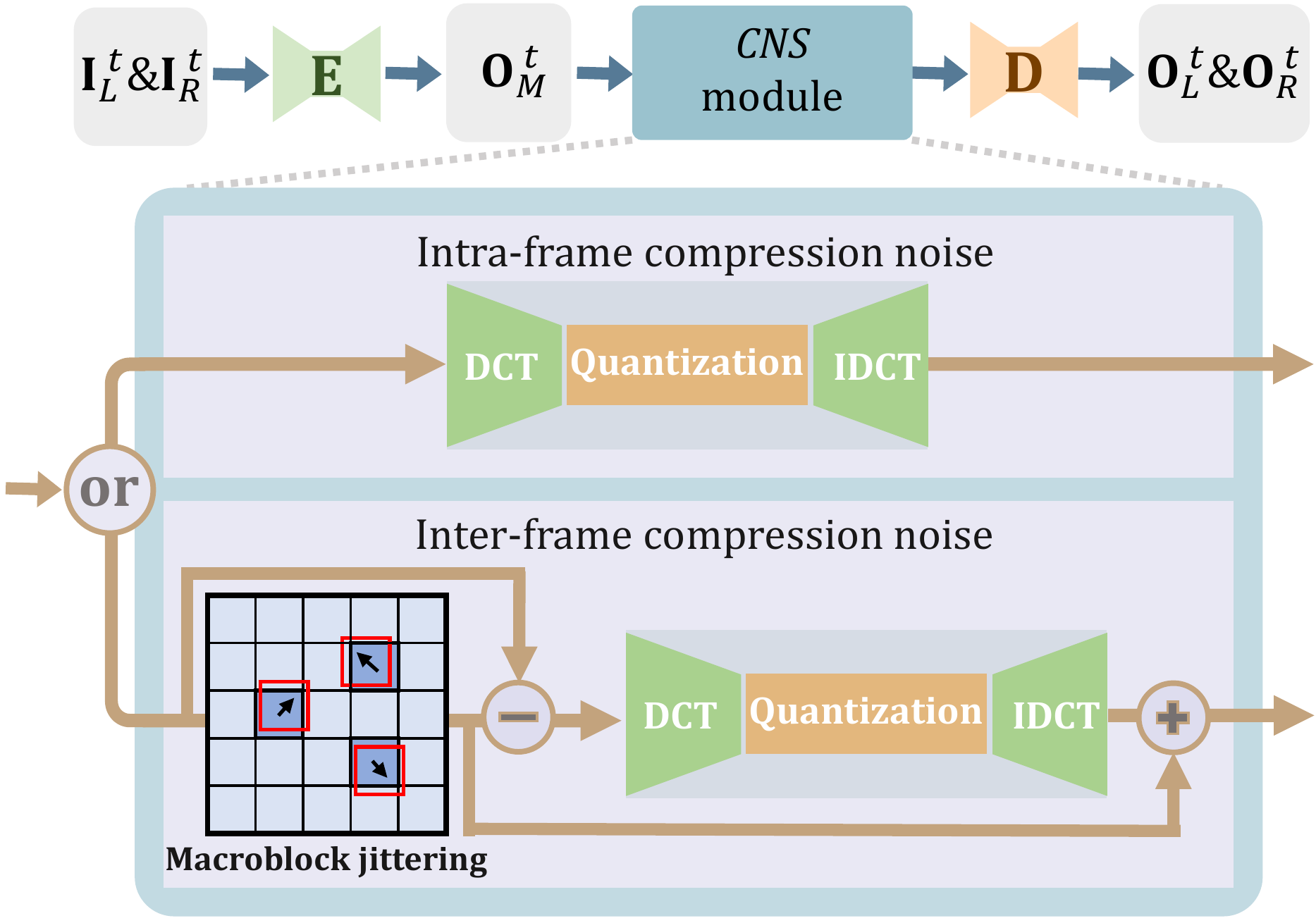}
	\vspace*{-1.5mm}
	\caption{
		Our extended framework with the \emph{compression noise simulation (CNS)} module (pipeline on top).
		The \emph{CNS} module between the encoder ($E$) and decoder ($D$) is designed to simulate the injection of intra- and inter-frame noise introduced by the video codec.
		DCT means Discrete Cosine Transform and IDCT is its inverse, while Macroblock means $16\times16$ pixels block.
	}
	\label{fig:model_exstension}
	\vspace*{-2mm}
\end{figure}


For distribution to users, mononized videos could be streamed by using lossy video codecs, e.g., H.264/MPEG4-AVC~\cite{wiegand2003overview}, H.265/HEVC~\cite{sullivan2012overview}, VP9~\cite{mukherjee2015technical}, and the newly issued AOMedia Video 1 (AV1)~\cite{chen2018overview}.
When streaming the mononized video at low bit-rates, the stereo information encoded in mononized videos may be distorted by the codecs, leading to bad restoration of the binocular video.

To resist the compression perturbation when collaborating with lossy video codecs, we further design an extended framework by inserting the \emph{compression noise simulation (CNS)} module (Figure~\ref{fig:model_exstension}) into our framework when we train the whole framework.
%
By introducing codec-like noise during the training, the framework can better learn to encode the stereo information in a compression-friendly mode, as well as to better restore the binocular frames from the distorted mononized frame.
We design the \emph{CNS} module to simulate the following two kinds of video codec noise:
\begin{itemize}
	%
	\item[(i)]
	intra-frame noise.
	We employ the DCT quantization~\cite{RobertsonS05} to simulate the lossy operation inside a frame; and
	\vspace*{1mm}
	\item[(ii)]
	inter-frame noise.
	This kind of noise mainly comes from \new{the quantization error of the residual signal due to the} inaccurate motion compensation in conventional video codecs; we simulate it by macroblock ($16\times16$) jittering, \new{as depicted in the lower half of Figure~\ref{fig:model_exstension}}.
	The jittering probability is up to the variance of the estimated optical flow.
\end{itemize}

During the network training, we randomly choose to simulate intra- or inter-frame noise (Figure~\ref{fig:model_exstension}), corresponding to the cases of I and P frames in conventional video codecs.
\section{Loss Function}
\label{sec:loss}
To drive the network training, we formulated three loss terms in the objective function (Eq.~\eqref{equation:lossForVideo}), namely the monocular loss, invertibility loss, and temporal loss.

\subsection{Monocular Loss}
\label{subsec:appearance_loss}

\begin{figure}[!t] 
	\centering
	\includegraphics[width=0.955\linewidth]{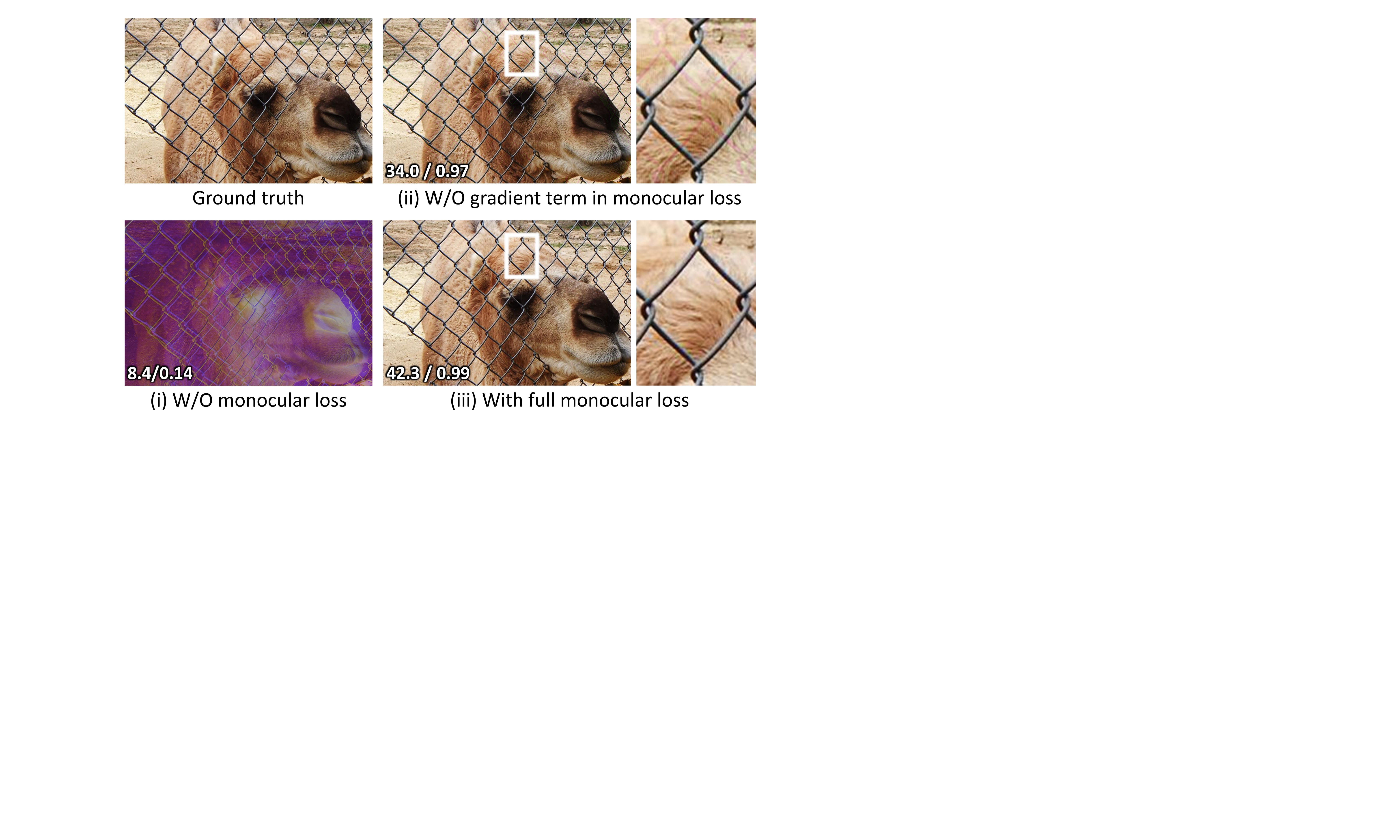}
	\caption{
		Effect of the monocular loss $\mathcal{L}_M$.
		(i)-(iii) Mononized frames $\mathbf{O}_M^t$ produced with three different forms of monocular loss.
		The numbers in each result show the corresponding PSNR and SSIM values.
	}
	\label{fig:mono_loss}
\end{figure} 

The monocular loss aims at producing an ordinary mononized frame $\mathbf{O}_M^t$ that looks like the input.
Without loss of generality, we choose the input left view $\mathbf{I}_L^t$ as the reference.
%
Therefore, we minimize the $L_2$ difference between $\mathbf{O}_M^t$ and $\mathbf{I}_L^t$.
However, as $\mathbf{O}_M^t$ contains the visual information from both $\mathbf{I}_L^t$ and $\mathbf{I}_R^t$, using the $L_2$ alone is insufficient and may result in noticeable patterns, which are closely related to the image details in $\mathbf{I}_R^t$ (Figure~\ref{fig:mono_loss} (top-right)).

To suppress the artifact, we further minimize the Charbonnier difference~\cite{charbonnier1994two} between the gradients of $\mathbf{O}_M^t$ and $\mathbf{I}_L^t$ in the monocular loss:
\begin{equation}\label{Eq:appearance_loss}
\mathcal{L}_M
=
\mathbb{E}_{\{\mathbf{I}_L^t, \mathbf{I}_R^t\} \in \mathcal{S}}\{
||\mathbf{O}_M^t-\mathbf{I}_L^t||_2
+
\alpha\cdot
\rho(\bigtriangledown\mathbf{O}_M^t-\bigtriangledown\mathbf{I}_L^t)\} \ ,
\end{equation}
where 
$\mathbb{E}$ denotes the average expectation over $N$ frames in a sequence\new{;}
$\alpha$ is a weight;
$\bigtriangledown$ is the gradient; \new{and}
$\rho(x) = \sqrt{x^2+\epsilon^2}$ is the Charbonnier $L_1$ function~\cite{charbonnier1994two}, where the constant $\epsilon$  is set to $1\times10^{-6}$.
Figure~\ref{fig:mono_loss} (bottom-right) shows an example restored frame produced with this monocular loss, in which the artifact has been greatly suppressed.

\subsection{Invertibility Loss}
\label{subsec:invertibility_loss}

The ability of restoring the binocular frames from mononized frame $\mathbf{O}_M^t$ is secured by the \textit{invertibility loss} $\mathcal{L}_I$. It minimizes the difference between the restored binocular frame pair $\{\mathbf{O}_L^t, \mathbf{O}_R^t\}$ and the original inputs $\{\mathbf{I}_L^t, \mathbf{I}_R^t\}$:
\begin{equation}\label{Eq:invertibility_loss}
\mathcal{L}_I = \mathbb{E}_{\{\mathbf{I}_L^t, \mathbf{I}_R^t\} \in \mathcal{S}}\{(1-\beta)\cdot||\mathbf{O}_L^t-\mathbf{I}_L^t||_2+\beta\cdot||\mathbf{O}_R^t-\mathbf{I}_R^t||_2\} \ ,
\end{equation}
where $\beta$ is a weight to balance the quality of the restored left and right views.
Restoring the right view is much harder than the left one, since the left view is taken as the reference for forming the mononized view.
Hence, we set $\beta$ to $0.99$ in practice.
Note that $\mathcal{L}_I$ effectively imposes constraints over the parameters in both the encoding and decoding networks, as we jointly train the two networks to produce $\{\mathbf{O}_L^t, \mathbf{O}_R^t\}$ in our framework.

\subsection{Temporal Loss}
\label{subsec:temporal_loss}

Mononizing binocular images can be achieved with the above two loss terms $\mathcal{L}_M$ and $\mathcal{L}_I$. 
However, using them alone is inadequate to ensure the temporal coherence 
when mononizing binocular videos.
As mentioned in Section~\ref{sec:overview}, we add recurrent connections in the framework to introduce the information of previous frame into the current one, but doing so cannot drive the model to learn how to utilize such information.
Hence, 
we design the temporal loss to maintain the coherence between successive frames:
\begin{eqnarray}\label{Eq:temporal_loss}
\mathcal{L}_T= \mathbb{E}_{\{\mathbf{I}_L^t, \mathbf{I}_R^t\} \in \mathcal{S}} \ \{ \ 
||\mathcal{W}(\mathbf{O}_M^{t-1}, \mathbf{F}_L^t)-\mathbf{O}_M^t||_2
\hspace*{3.25mm} \\ \nonumber
+ \
||\mathcal{W}(\mathbf{O}_L^{t-1}, \mathbf{F}_L^t)-\mathbf{O}_L^t||_2
\hspace*{4.20mm} \\ \nonumber
+ \ \gamma \cdot \
||\mathcal{W}(\mathbf{O}_R^{t-1}, \mathbf{F}_R^t)-\mathbf{O}_R^t||_2 \ \} \ ,
\end{eqnarray}
where 
$\mathbf{F}_L^t$ is the estimated optical flow from $\mathbf{I}_L^{t-1}$ to $\mathbf{I}_L^t$;
$\mathbf{F}_R^t$ is the estimated optical flow from $\mathbf{I}_R^{t-1}$ to $\mathbf{I}_R^t$; 
$\mathcal{W}(\mathbf{X},\mathbf{F})$ produces a warped image of $\mathbf{X}$ by using the optical flow $\mathbf{F}$; and
$\gamma$ is a weight.
Since restoring a high-quality $\mathbf{O}_R^t$ is more challenging than $\mathbf{O}_L^t$, we put a relatively larger weight on the term for $\mathbf{O}_R^t$.
%
Comprehensive qualitative and quantitative analysis will be presented in Section~\ref{subsubsec:temporal} to demonstrate the effect of the temporal loss.


\section{Training}
\label{sec:training}


\paragraph{Training data}
Publicly-available datasets with binocular frame sequences such as~\textit{KITTI}~\cite{GeigerL13} are often too specific in genre.
Hence, we compile a \textit{3D movie} dataset that contains
%
122 3D movie sequences with 720p resolution ($5,876$ frames in total) collected from~\textit{Inria}\footnote{Inria: \url{https://www.di.ens.fr/willow/research/stereoseg/}}~\cite{SeguinA15} and~\textit{YouTube}\footnote{YouTube: \url{https://www.youtube.com/}}.
Since some stereoscopic videos in YouTube were artificially produced by a naive mono to stereo conversion, we intentionally avoided them by estimating the disparity of each collected video and ignoring those with unnatural disparity.
Overall, the dataset covers eight types of scenes, e.g,~\textit{animals and pets},~\textit{architecture},~\textit{cartoon}, and~\textit{natural scenery}; see Supplemental material Section 2 for the detailed description.
Further, we employed the pre-trained~\textit{PWC-Net}~\cite{SunY18} to estimate the optical flows between consecutive frame pairs in each movie sequence.
Lastly, we randomly selected 69 sequences as the training set and used the remaining 53 sequences as the test set.

Besides, we employ a stereo image dataset, \textit{Flickr1024}~\cite{flickr1024}, which contains 1,024 binocular images of various categories, to train the image version of our method for comparison with other methods.
%
%
Here, we follow the official train/test split in Flickr1024.
%

\vspace*{-5pt}
\paragraph{Training details}
We implemented our encoding and decoding networks using PyTorch~\cite{PaszkeGMLBCKLGA19} and trained them jointly with the loss function defined in Eq.~\eqref{equation:lossForVideo}.
%
%
Each mini-batch training samples contains $N \times B$ frames, where 
$N$ is the number of consecutive frames; 
$B$ is the number of instances; and 
each frame is randomly cropped into $256\times256$ resolution during the training.
%
%
We empirically set $N = 4$ and $B = 16$ in the training.

We optimized our model by the Adam solver~\cite{KingmaB14}, in which the learning rate was initially set to $0.0001$ and further reduced by a factor of $3.33$ when the loss plateaus (known as ReduceLRonPlateau in PyTorch).
For the image version of our model, we trained the networks on the training set of \textit{Flickr1024} for $200$ epochs, while for the video version, we trained the networks for $300$ epochs on the training set of the compiled \textit{3D movie} dataset.
During the training, the video version of our framework was initialized by the image version of our trained framework model, since we empirically found that doing so improves the overall performance than simply initializing the network parameters from scratch.
The hyper-parameters in the loss function are set as following:
$\lambda_1 = 1.7$;
$\lambda_2 = 1.3$;
$\alpha = 0.1$;
$\beta = 0.99$; and
$\gamma = 10.0$.
\new{\emph{Code is available at the following GitHub page: \url{https://github.com/wbhu/Mono3D}.}
}

%

\newcommand{\Mimg}{$\mathbf{Mono3D}_\text{\footnotesize{img}}$}
\newcommand{\M}{$\mathbf{Mono3D}_\text{\footnotesize{video}}$}
\newcommand{\MwoPDF}{$\mathbf{Mono3D}_\text{\footnotesize{video}}^\text{\footnotesize{w/o}\tiny{\;PDF}}$}
\newcommand{\Mplus}{$\mathbf{Mono3D}_\text{\footnotesize{video}}\hspace*{-0.3mm}\texttt{\footnotesize ++}$}
\newcommand{\MwoCNS}{$\mathbf{Mono3D}_\text{\footnotesize{video}}^\text{\footnotesize{w/o}\tiny{\;CNS}}$}

\newcommand{\DS}{\emph{DeepSteno}}
\newcommand{\IG}{\emph{InvertGray}}
\newcommand{\KB}{\emph{3D Ken burns}}

\begin{figure*}[!t] 
	\centering
	\includegraphics[width=0.99\linewidth]{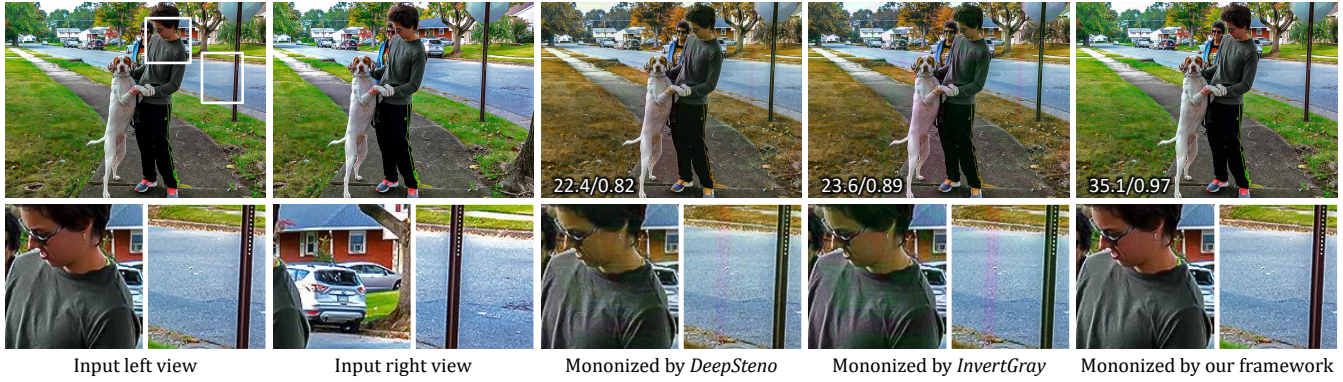}
	\vspace*{-2mm}
	\caption{
		From left to right: the input left and right views, followed by the mononized images produced by 
		\DS~(from~\cite{baluja2017hiding}),
		\IG~(from~\cite{XiaL18}),
		and the image version of our framework (\Mimg).
		Note the PSNR/SSIM values shown in each result.
		Our mononized result does not have obvious artifacts, such as color shifting and traces of objects that come from the right view; see the blown-up views on the bottom row.
	}
	\label{fig:compare_mono}
\end{figure*} 

\section{Results and Discussion}
\label{sec:results}


\subsection{Qualitative Evaluation}
\label{subsec:qualitative_evaluation}

Figure~\ref{fig:showcase} at the end of this paper showcases four example results produced by our method, featuring indoor and outdoor contents with close-up objects and faraway scenery.
%
In each example, we show a tabular figure of 2$\times$4 images: the input left and right views (1st column);
mononized view and its difference from the input left view (2nd column);
restored left and right views (3rd column); and
their differences from the corresponding inputs (4th column).
The numbers in each result (\OM, \OL, and \OR) show the PSNR and SSIM values compared with the corresponding input (ground truth).
For the difference maps, we compute the absolute pixel value difference in the scale of [0,255] and color-code the difference value. The mean absolute differences are often very small (only around two).

In Figure~\ref{fig:showcase}, the top three examples are still pictures, whereas the bottom-most one is a video frame in the test dataset (more video results can be found in the supplemental video).
Since video examples are usually less challenging due to small disparity, we pick three more challenging still pictures to show the capability of our network to handle occlusions and large disparity.
See particularly the top example, the baby monkey on top of the right view (\IR) is mostly occluded by the front monkey in the left view (\IL), so it is visually absent in the mononized image (\OM); {\em yet, our method can restore it (\OR) solely from the mononized image (\OM)}, just like the face of the little girl shown in Figure~\ref{fig:teaser}.
Also, the pixel value differences in the occluded region, e.g., the baby monkey, are not obviously higher, as revealed in the difference map.
These results demonstrate the capability of our method to implicitly encode the stereo information in a nearly-imperceptible form inside the mononized view and later restore from it the binocular views.
More visual comparison results can be found in Supplemental material Section 4.


\subsection{Comparison with Other Methods}
\label{subsec:compare_sota}

So far, no methods have been developed for mononizing binocular images and videos.
Hence, to evaluate and demonstrate the quality of our method, we adopt the following three related works for comparison:
(i) deep stenography~\cite{baluja2017hiding} (denoted as \DS), in which we take the right view as the secret image and use its preparing and hiding networks to conceal the right view in the left view, then further reconstruct the right view from the stenographed image by its reveal network;
(ii) the reversible framework in~\cite{XiaL18} (denoted as \IG), in which we concatenate the left and right views along the channel dimension, feed the result into its encoding network to produce the mononized view, then use its decoding network to restore the left and right views; and
(iii) further, we explore the possibility of dropping the right view and synthesizing it from the left one using the recent novel-view-synthesis method, \KB~\cite{NiklausMYL19}.
Clearly, view synthesis might not produce high-quality results; here, we take it as a baseline to see if our method can encode stereo information in the mononized (left) view for reconstructing a better right view.

Since the three methods are originally designed for still pictures, for a fair comparison, we adopted our framework for mononizing binocular images (denoted as \Mimg) without the recurrent connections, temporal loss, and CNS module, and trained it on the \textit{Flickr1024} dataset (Section~\ref{sec:training}).
%
For \DS, we implemented its method according to its paper, as there is no public code.
For \IG, we obtained code from its project webpage.
Then, we re-trained their networks on \textit{Flickr1024} with our loss function, since their original loss functions are not designed for mononizing binocular images.
%
%
For \KB, we adopted their released trained model and manually tuned the camera pose to best align the synthesized right view with the ground-truth right view.

\subsubsection{Evaluation on the mononized views}

%
Figure~\ref{fig:compare_mono} shows the mononized views produced by \DS, \IG, and our framework, in which we can see color shifting problem in the results of \DS~and \IG.
Also, we can observe obvious traces of objects from the right view; see the blown-up views on the bottom of Figure~\ref{fig:compare_mono}.
Compared with our framework, \DS~does not take into account the relations between the container (left view) and secret (right view) images, whereas \IG~cannot be directly extended for mononizing binocular images.
\IG~is designed to handle aligned luminance and chrominance in the invertible grayscale problem, so it cannot effectively harvest the long-range correspondences between the left and right views.
Besides, both cannot maintain frame-to-frame coherence in the video inputs.
Thanks to the architecture and the pyramidal deformable fusion module, our framework can learn to leverage the correspondences between left and right views to mononize binocular images.
From Figure~\ref{fig:compare_mono}, we can see that our mononized view does not exhibit obvious traces from the right view, meaning that the stereo information can be implicitly encoded in a nearly-imperceptible form.
%

Further, we quantitatively compare the visual quality of the mononized views produced by various methods on the whole \textit{Flickr1024} test set in terms of PSNR and SSIM.
%
%
%
As shown in the Mono-view column of Table~\ref{tab:compare_sota}, the average PSNR and SSIM values of our mononized views (37.8 \& 0.97) are far higher than those of \DS~(26.1 \& 0.81) and \IG~(28.0 \& 0.81).
These statistical results quantitatively demonstrate the effectiveness of our method.

\begin{figure*}[!t] 
	\centering
	\includegraphics[width=0.99\linewidth]{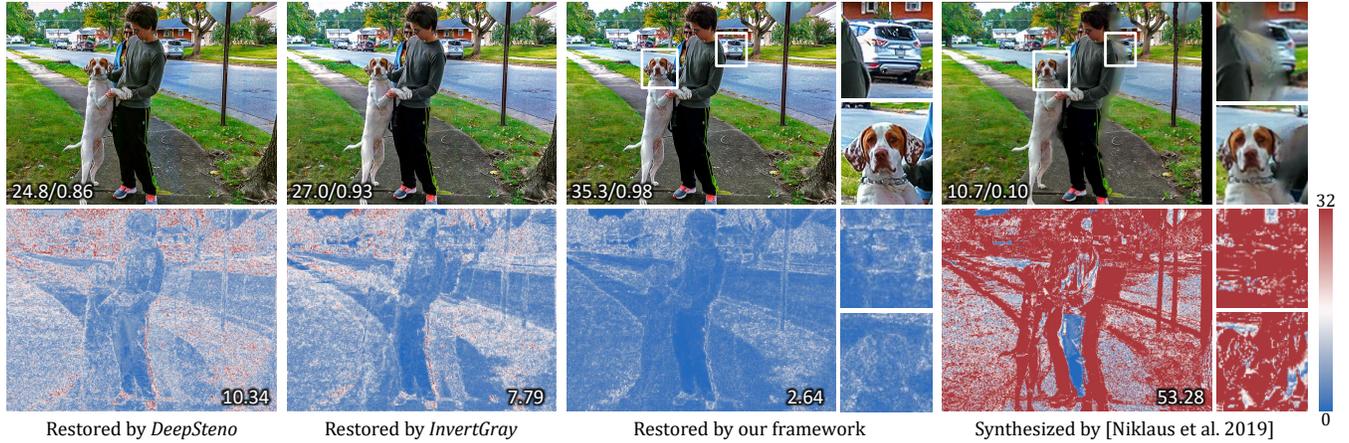}
	\vspace*{-2mm}
	\caption{
		Restored/synthesized right views produced by \DS, \IG, the image version of our framework (\Mimg), and the novel-view-synthesis method~\cite{NiklausMYL19}.
		PSNR/SSIM values and mean absolute error (in the scale of 255) are shown in each result.
		We can see from the error maps on the bottom row that our restored view is significantly \new{closer} to the ground-truth results with very small pixel color differences.
	}
	\vspace*{-3mm}
	\label{fig:compare_right}
\end{figure*} 

\begin{table}[!t]
	\centering
	\renewcommand{\tabcolsep}{4pt}
	\caption{
		Visual quality of the mononized views and restored binocular (left \& right) views produced by 
		\DS, \IG, 
		and our framework (\Mimg) on \textit{Flickr1024}.
		Note that \DS{} does not restore the left view,
		as its reveal network is only designed for restoring the secret image.
	}
	\vspace{-2mm}
	\begin{tabular}{c||cc|cc|cc}
		\hline
		\multirow{2}{*}{Methods} & \multicolumn{2}{|c|}{Mono-view} & \multicolumn{2}{|c|}{L. Bino-view} & \multicolumn{2}{|c}{R. Bino-view} \\ \cline{2-7}
		&     PSNR      &      SSIM       &     PSNR      &        SSIM        &     PSNR      &       SSIM        \\ \hline
		\DS &     26.1      &      0.81       &     \----     &       \----        &     27.9      &       0.88        \\
		\IG &     28.0      &      0.89       &     28.7      &        0.92        &     30.7      &       0.92        \\
		\Mimg        & \textbf{37.8} &  \textbf{0.97}  & \textbf{38.3} &   \textbf{0.99}    & \textbf{37.3} &   \textbf{0.98}   \\ \hline
	\end{tabular}
	\label{tab:compare_sota}
	\vspace{-2mm}
\end{table}

\subsubsection{Evaluation on restored/synthesized views}
Next, we compare the restored right view produced by \DS, \IG, and our framework.
From the first three columns shown in Figure~\ref{fig:compare_right}, we can see that our result has much higher PSNR and SSIM, whereas those of \DS~and \IG~contain traces of objects from the left view, as can be seen in the error maps shown in the figure.
Overall, the average PSNR of our restored right views are well above 35dB, compared with those of \DS~and \IG.
The statistical results for the restored right views in Table~\ref{tab:compare_sota} further confirm the findings.
Note that similar statistical results are also obtained for the restored left views (Table~\ref{tab:compare_sota}).
Here, we do not consider the restored left view when comparing with~\DS, since its reveal network is designed only for restoring the secret image, so it restores only the right view but not the left one from the stenographed image.

%
%

Further, the last column in Figure~\ref{fig:compare_right} shows the synthesized right view by \KB.
As discussed earlier, it is very hard to estimate accurate depth and infer plausible results for the occluded regions, so the results of \KB~tend to be blurry on the inpainted regions.
%
%
We admit that this comparison is not entirely fair, as inferring and restoration are not directly comparable. 
Yet, the comparison gives evidence that our encoding-and-decoding approach is able to recover the stereo information not available in the left view.

%


\newcommand{\Msingle}{$\mathbf{Mono3D}_\text{\footnotesize{video}}^\text{\footnotesize{single-scale}}$}
\newcommand{\MwoDConv}{$\mathbf{Mono3D}_\text{\footnotesize{video}}^\text{\footnotesize{w/o}\tiny{\;DConv}}$}

\subsection{Quantitative Evaluation}
\label{subsec:quantitative_evaluation}

Next, we quantitatively evaluate our method on the test set of the \textit{3D movie} dataset (Section~\ref{sec:training}).
To verify the effectiveness of some key designs in our method, we consider \new{five} variants of our method:
\begin{itemize}
	\item
	\M: our full method for mononizing binocular videos;
	\new{
	\item
	\Msingle: we remove the first two levels of pyramidal features in \M{} and use only the third-level feature;
	\item 
	\MwoDConv: from \M, we replace the deformable convolution in the PDF modules with conventional convolution;
	}
	\item
	\MwoPDF: from \M, we remove the PDF modules, directly concatenate \PLt{} and \PRt{} into \PMt{} in the encoding network, and separate \PMtildet{} into \PLtildet{} and \PRtildet~simply along the channel dimension in the decoding network; and
	%
	\item
	\Mimg: our method for mononizing binocular images, i.e., \M{} without the recurrent connections from previous frames, temporal loss, and CNS module (Figure~\ref{fig:overview}).
\end{itemize}

\subsubsection{Evaluation on frame quality}
\label{subsubsec:frame_quality}

\begin{table}[!t]
	\centering
	\renewcommand{\tabcolsep}{4pt}
	\caption{
		Frame quality of the mononized and restored binocular views produced by \new{the five} variants of our method (\Mimg, \new{\Msingle, \MwoDConv,} \MwoPDF, and \M) over the entire test set of \textit{3D movie}.
	}
	\vspace{-2mm}
	\begin{tabular}{l||cc|cc|cc}
		\hline
		\multirow{2}{*}{Method variants} & \multicolumn{2}{|c|}{Mono-frame} & \multicolumn{2}{|c|}{L. Bino-frame} & \multicolumn{2}{|c}{R. Bino-frame} \\ \cline{2-7}
		                                 &     PSNR      &       SSIM       &     PSNR      &        SSIM         &     PSNR      &        SSIM        \\ \hline
		\Mimg                            &     38.6      &       0.97       &     39.5      &        0.98         &     37.9      &        0.97        \\
		\new{\Msingle}                   &  \new{29.7}   &    \new{0.90}    &  \new{30.1}   &     \new{0.92}      &  \new{30.5}   &     \new{0.92}     \\
		\new{\MwoDConv}                  &  \new{36.6}   &    \new{0.94}    &  \new{36.8}   &     \new{0.95}      &  \new{35.7}   &     \new{0.93}     \\
		\MwoPDF                          &     34.1      &       0.92       &     34.9      &        0.94         &     33.9      &        0.93        \\
		\M                               & \textbf{39.0} &  \textbf{0.98}   & \textbf{39.8} &    \textbf{0.99}    & \textbf{38.7} &   \textbf{0.98}    \\ \hline
	\end{tabular}
	\label{tab:quantitative_results}
	\vspace{-2mm}
\end{table}

We adopt PSNR and SSIM to measure the frame quality of our results relative to the inputs as the ground truths.
Table~\ref{tab:quantitative_results} lists the PSNR and SSIM values of the mononized and restored binocular frames produced by the \new{five} variants of our method on the entire \textit{3D movie} test set.

Comparing the statistical results of \Mimg{} and \M, we can see that both can produce high-quality results with PSNR well above 35dB, while \M{} performs slightly better, showing that 
the recurrent connections help introduce extra useful information from previous time frames to improve the performance.
Note that PSNR and SSIM only measure the quality of individual frames without considering the temporal quality.
More analysis on the temporal frame quality will be presented in Section~\ref{subsubsec:temporal}.

\new{
Comparing the results of \Msingle{} and \M{} as shown in Table~\ref{tab:quantitative_results}, we can see that \M{} performs significantly better.
Since the pyramidal features help harvest both low- and high-level information, \M{} leads to better results for both the mononized and restored binocular frames.
}

\new{
Comparing the results of \MwoPDF{} and \MwoDConv{}, we can see that \MwoDConv{} performs better, since the pyramidal fusion improves the encoding and decoding of stereo information to and from the mononized frames.
More importantly, comparing the statistical results of \MwoDConv{} and \M, we can see that \M{} performs even better.
%
%
Such result quantitatively shows the effectiveness of the PDF module for exploiting long-range correspondences and fusing features from the left and right views.
}

\subsubsection{Evaluation on temporal coherence}
\label{subsubsec:temporal}

\begin{figure}[!t]
	\centering
	\includegraphics[width=0.99\linewidth]{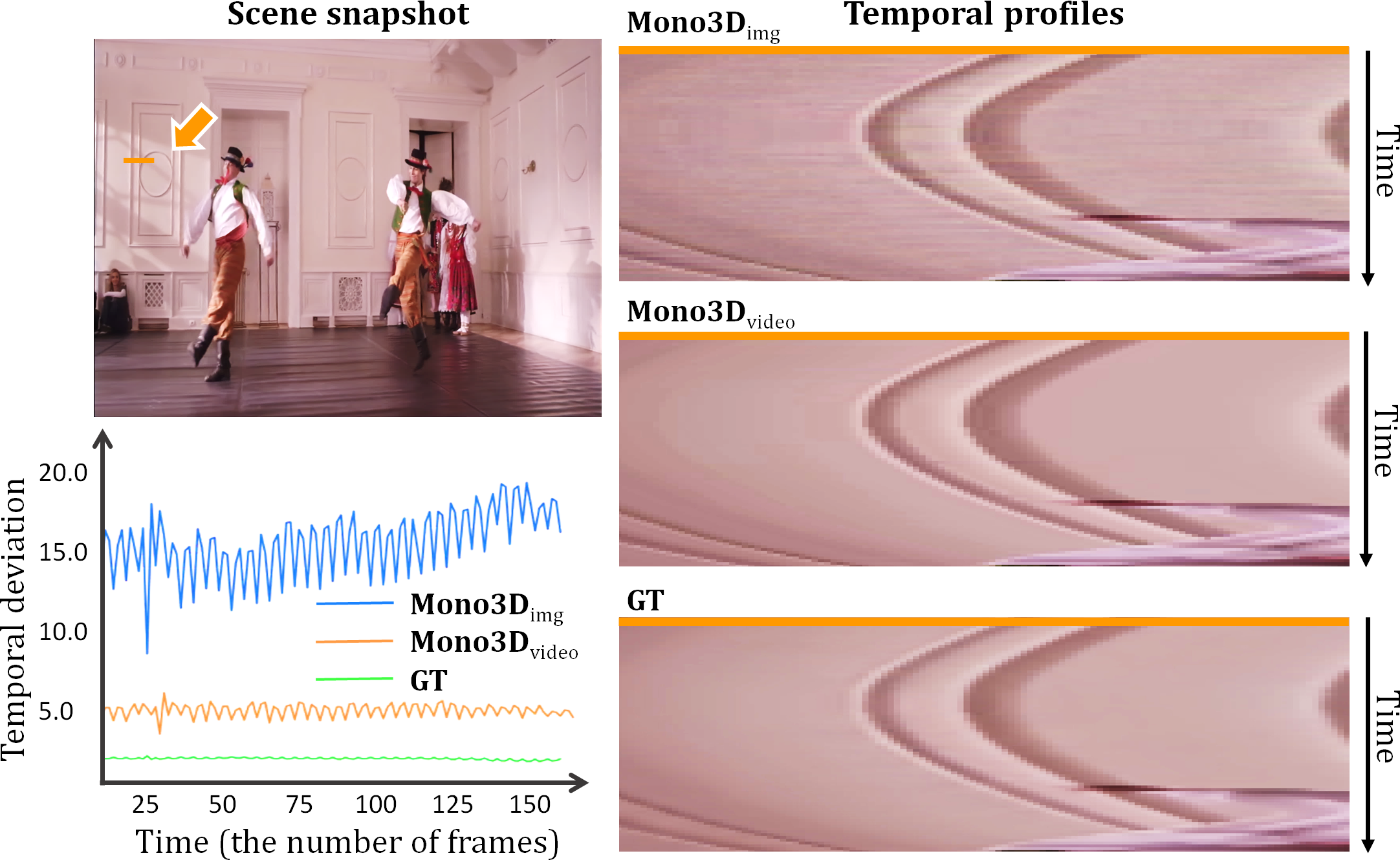}
	\vspace*{-2mm}
	\caption{Temporal coherence evaluation.
		The top-left image shows a snapshot of the scene;
		the right three images 
		show the temporal profiles of \Mimg, \M, and $\mathbf{GT}$ (ground truth) along the pixels in the orange line segment marked by the orange arrow in the scene snapshot; and
		the bottom-left plot shows the temporal deviation of the right binocular frames restored by \Mimg{} and \M, and of the raw frames directly from the ground truth, i.e., the input binocular video.
	}
	\label{fig:temporal_result}
\end{figure}

There is no standard way to measure the temporal coherence of a video.
The most direct way is to show the video to humans and let them evaluate the temporal coherence.
%
Readers are recommended to watch our supplement video for the evaluation.
Besides, we extract a line of pixels at a fixed location in videos, and stack them over time as an image of {\em temporal profile\/}; see the right-hand side of Figure~\ref{fig:temporal_result} for examples.
Comparing the temporal profiles of the restored right frames from \Mimg{} and \M, as well as the raw frames from the ground-truth input (\textbf{GT}), we can see that the temporal profile of \M{} is much smoother than that of \Mimg{}, and it also looks more similar to the temporal profile of the ground truth.

To quantitatively evaluate the temporal coherence, we explore an observation that if a generated video (\OMt, \OLt, or \ORt) is temporally coherent, each frame in the video should be more predictable from the previous one via optical flow estimated between frames in the original video.
Based on this idea, we first use~\textit{PWC-Net}~\cite{SunY18} to estimate the optical flows between each pair of successive frames in the input, i.e., $\mathbf{F}_x^t$ from $\mathbf{I}_x^{t-1}$ to $\mathbf{I}_x^t$, where $x$ is $L$ or $R$ for left or right view, respectively.
Then, we warp each frame in the generated videos, and compute the {\em warping deviation\/} between the warped frame ($\mathcal{W}(\mathbf{O}_x^{t-1},\mathbf{F}_x^t)$) and next frame ($\mathbf{O}_x^t$) in the video:
\begin{equation}
\label{eq:delta_O}
\Delta_x^t \ = \ | \ \mathcal{W}(\mathbf{O}_x^{t-1},\mathbf{F}_x^t) \ - \ \mathbf{O}_x^t \ | \ ,
\end{equation}
where $x$ is $M$, $L$, or $R$, and $\mathbf{F}_M^t$ is taken as $\mathbf{F}_L^t$.
Likewise, we compute also the warping deviation for the input binocular videos:
\begin{equation}
\label{eq:delta_I}
\hat{\Delta}_x^t \ = \ | \ \mathcal{W}(\mathbf{I}_x^{t-1},\mathbf{F}_x^t) \ - \ \mathbf{I}_x^t \ | \ ,
\end{equation}
where $x$ is $L$ or $R$, and $\hat{\Delta}_x^t$ are mostly zeros, except in areas that $\mathbf{I}_x^t$ cannot be warped from $\mathbf{I}_x^{t-1}$, e.g., occlusion and lighting changes.
For temporally-coherent videos, $\Delta_x^t$ should be coherent with $\hat{\Delta}_x^t$.
Hence, we define the {\em temporal deviation\/} of a generated video as the mean absolute relative error of $\Delta_x^t$ w.r.t. $\hat{\Delta}_x^t$ over the whole image:
%
\begin{equation}
\label{eq:temp error}
\new{
\sigma^t \ = \ (\prod_{i=1}^{W} \prod_{j=1}^{H} \prod_{c=1}^{3} {| \frac{{\Delta_x^t}_{(i,j,c)}}{\hat{{\Delta}_x^t}_{(i,j,c)} + \epsilon} |})^{\frac{1}{H \times W \times 3}},
}
\end{equation}
where $\sigma^t$ is a scalar that indicates the temporal deviation at time $t$,
\new{$W$ is image width, $H$ is image height, $(i, j)$ is pixel index, $c$ is RGB channel index, and}
$\epsilon$ is set as $1\times10^{-3}$ in practice to avoid division by zero.
Note that $\sigma^t \approx 1$ shows high temporal coherence, and vice versa.
It is because if a given video is not temporally coherent, its warping deviation ($\Delta_x^t$) will be very far from that of the input ($\hat{\Delta}_x^t$).

\begin{table}[!t]
	\centering
	\renewcommand{\tabcolsep}{4pt}
	\caption{Temporal coherence evaluation via temporal deviation (Eq.~\eqref{eq:temp error}) on the \textit{3D movie} test set.
		Better temporal coherence should be closer to one.}
	\vspace{-2mm}
	\begin{tabular}{l||c|c|c}
		\hline
		Method variants & Mono-frame & L. Bino-frame & R. Bino-frame \\ \hline
		\Mimg           & \new{1.90} &  \new{1.55}   &  \new{4.73}   \\
		\M              & \new{1.24} &  \new{1.10}   &  \new{1.22}   \\ \hline
	\end{tabular}
	\label{tab:temporal_metric}
	\vspace{-2mm}
\end{table}

Figure~\ref{fig:temporal_result} presents plots of $\sigma^t$ over time for the case of \M{} and \Mimg, as well as for the ground truth (\textbf{GT}), typically for the right binocular frames in a challenging video example.
Here, if we compare the warping deviations of \textbf{GT} with itself, the resulting $\sigma^t$ values are almost one, as shown in the plot for \textbf{GT}.
More importantly, we can see that the plot of \M{} is always closer to the plot of \textbf{GT} and lower than the plot of \Mimg, thus showing that \M{} produces more temporally-coherent videos than \Mimg.
To statistically confirm the results,
we compute the \new{geometric} mean $\sigma$ values over time for all the videos in the \textit{3D movie} test set.
From the \new{geometric} mean $\sigma$ values shown in Table~\ref{tab:temporal_metric}, we can see that the temporal deviation of all the results produced by \M{} are very close to one, compared with \Mimg, thus demonstrating how temporal coherence is improved by having the temporal loss and recurrent connections in our framework.

\subsubsection{Timing performance}

We implemented our method using \textit{PyTorch} and ran all experiments on a PC equipped with a Titan Xp GPU and Intel Core i9-7900X@3.30GHz CPU.
We evaluated the time performance of our method for one frame in multiple resolutions.
Table~\ref{tab:efficiency} shows the timing statistics, where we exclude the time to transfer data between the GPU and CPU, since it has nothing to do with the method and can be optimized using memory cache and pipeline techniques.
From the results, we can see that our method performance can support real-time applications on \textit{1080p} videos.
\begin{table}[!t]
	\centering
	\renewcommand{\tabcolsep}{6pt}
	\caption{Timing statistics of our method for one frame (in milliseconds).}
	\vspace{-2mm}
	\begin{tabular}{c||c|c}
		\hline
		    Resolution     & Encoding & Decoding \\ \hline
		 $480 \times 720$  &   10.9   &   10.5   \\ \hline
		$720 \times 1280$  &   18.6   &   18.5   \\ \hline
		$1080 \times 1920$ &   27.3   &   26.9   \\ \hline
	\end{tabular}
	\label{tab:efficiency}
	\vspace{-1mm}
\end{table}

\begin{figure*}[!t]
	\centering
	\includegraphics[width=0.99\linewidth]{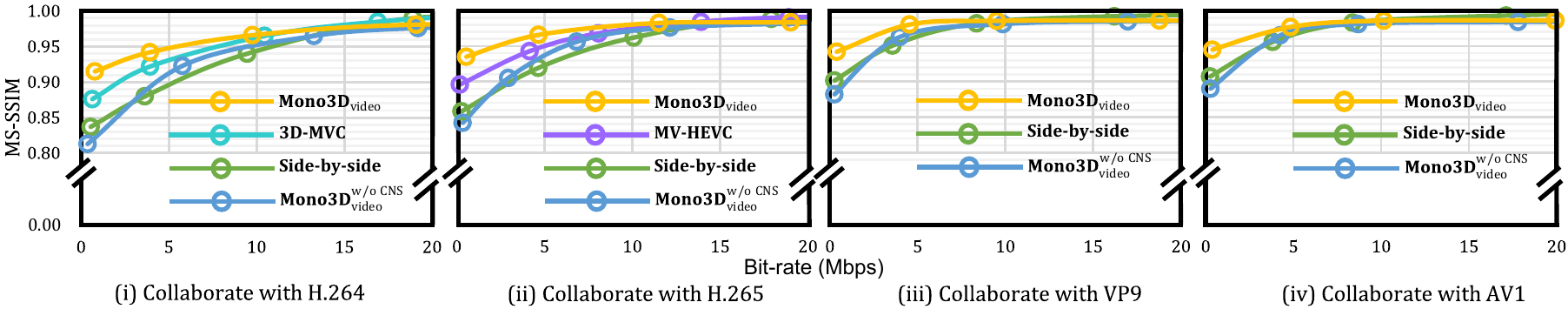}
	\vspace{-2mm}
	\caption{Rate-distortion (R-D) curves of \M, \textbf{3D-MVC}, \textbf{Side-by-side}, and \MwoCNS.
		The distortion is measured by MS-SSIM and bit-rate is represented as megabits per second (Mbps).
		All the restored binocular videos have $1280\times720$ resolution and 25 FPS for both the left and right views.
}
	%
	\label{fig:compression_analysis} 
	\vspace{-2mm}
\end{figure*}

\subsection{Compatibility with Video Codecs}
\label{subsec:video_codecs}


Next, we evaluate the compatibility of our framework with common video codecs:
(i) H.264/MPEG4-AVC~\cite{wiegand2003overview} and (ii) H.265/HEVC~\cite{sullivan2012overview}, 
(iii) VP9~\cite{mukherjee2015technical}, which is commonly-used in web browsers, and 
(iv) AOMedia Video 1 (AV1)~\cite{chen2018overview}, which aims to be the next-generation codec.
Overall, we passed the mononized videos produced by our framework to these codecs for encoding and decoding, thus 
emulating the use of our mononized videos as regular monocular videos in common video platforms.
Then, we fed the mononized videos decoded by the codecs into our decoder network to restore the binocular videos, and measured the quality of the restored binocular videos relative to the originals.

To the best of our knowledge, there is no multi-view codec extension that is backward compatible with multiple monocular video codecs, while our method is fully backward-compatible.
So far, there are only codec-specific multi-view extensions as discussed in Section~\ref{sec:codec}.
Here, our comparison includes the following methods:
\begin{itemize}
	\item
	\M: our full method for mononizing binocular videos;
	\item
	\MwoCNS: our \M{} without the CNS module;
	\item
	\textbf{Side-by-side}: concatenate each pair of left and right frames side-by-side~\cite{vetro2010frame}, and encode them using each of the four video codecs mentioned above;
	\item
	\textbf{3D-MVC}\footnote{Available at~\url{https://www.videohelp.com/software/FRIM}}: an multi-view extension of H.264/MPEG4-AVC; and
	\item
	\textbf{MV-HEVC}\footnote{Available at~\url{https://github.com/listenlink/3D-HEVC}}: an multi-view extension of H.265/HEVC.
\end{itemize}
In this experiment, we followed existing video compression research to use MS-SSIM~\cite{wang2003multiscale} to measure the quality of the restored video over the test dataset, and plotted the rate-distortion (R-D) curves (Figure~\ref{fig:compression_analysis}) to explore video quality vs. bit-rate.


Comparing the R-D curves of \textbf{Side-by-side}, \textbf{3D-MVC}, \textbf{MV-HEVC} and our \M in Figure~\ref{fig:compression_analysis}, we can see that \M{} outperforms all others when encoding with low bit-rates (< 10 Mbps) for all the four codecs.
Importantly, the multi-view extensions, e.g., \textbf{3D-MVC} and \textbf{MV-HEVC}, all depend on the monocular codecs (for instance, we cannot apply MV-HEVC to the H.264/MPEG4-AVC codec), the \textbf{Side-by-side} method is not compatible with existing monocular TVs, whereas our \M{} is fully compatible with the existing monocular codecs and TV systems.
When encoding with high bit-rates (> 15 Mbps), we can see that \M{} performs slightly worse than \textbf{3D-MVC}, \textbf{MV-HEVC} and \textbf{Side-by-side}.
Yet, the overall video quality is very close and comparable with others, as shown in the plots, for all the four codecs.


Comparing the R-D curves of \M and \MwoCNS{}, we can see that \M{} significantly outperforms \MwoCNS{} at low bit-rates for all the four codecs.
These results show that our CNS module helps the framework to learn to encode the stereo information in a compression-friendly manner by introducing codec-like noise in the network training.
Also, \M{} and \MwoCNS{} perform similarly at high bit-rates, since the compression perturbation of modern codecs becomes too trivial at high bit-rates (with more storage resource in the encoding).
Moreover, the R-D curve trends of \M{} are similar for all the four plots, showing that our method is not sensitive to specific codecs.


Overall, the experimental results show the backward-compatibility of our framework with various common video codecs.
We can encode and stream our mononized videos on existing monocular platforms, just as the regular monocular videos.
On top of this, we can restore binocular videos from the streamed mononized videos for stereoscopic viewing, if a 3D display is available.


\subsection{User Study}
\label{subsec:user_study}

\begin{figure*}[!t]
\centering
\includegraphics[width=0.99\linewidth]{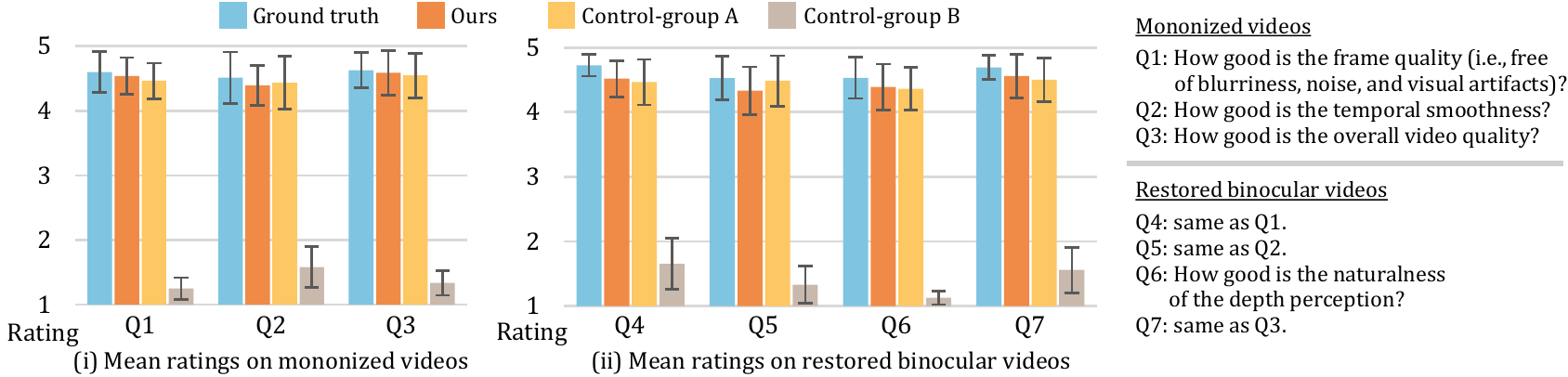}
\vspace*{-3mm}
\caption{User study results.
Mean ratings given by the participants on the mononized videos (i) and restored binocular videos (ii) for the cases of 
ground truth, our framework, control-group A, and control-group B (with slight noise, temporal flicker, etc.).
The error bar on each column indicates the standard deviation.
}
%
	\label{fig:userStudy}
	\vspace*{-2mm}
\end{figure*}

Further, to emulate the use of our mononized videos in existing video platforms, we conducted a user study to evaluate the perceptual quality of the mononized videos and restored binocular videos produced by \M, in \new{combination} with the H.264 video codec.
Here, we chose H.264, since our method has slightly lower performance with H.264 (Figure~\ref{fig:compression_analysis}).
Also, H.264 is the most common codec nowadays.
%
%
For the mononized videos, we evaluate the frame quality, temporal smoothness, and overall video quality.
For the restored binocular videos, besides the above three quantities, we additionally evaluate the depth perception of the binocular videos; see questions Q1-Q7 listed in Figure~\ref{fig:userStudy} (right).


\vspace{-3pt}
\paragraph{Preparation}
To start, we prepared four types of videos:
\begin{itemize}
	\vspace*{-1mm}
	\item[(i)] \emph{Ground truth}. We randomly selected eight video samples ($720p$, $25$ FPS), one per category from the \textit{3D movie} test set, as the ground-truth binocular videos, and simply regarded the left views as the ground-truth monocular videos;
	\item[(ii)] \emph{Ours}. From these ground truths, we generated the mononized videos using \M{}, then encoded and decoded each mononized video by the H.264 codec at $\sim$5 Mbps\footnote{The recommended bit-rate of YouTube is 5 Mbps for 720p 25 FPS monocular videos,~\url{https://support.google.com/youtube/answer/1722171}.}, and finally restored from them the binocular videos using \M{}.
%
	\item[(iii)] \emph{Control-group A}. We encoded and decoded the ground truths by the multi-view extension of H.264, 3D-MVC, at $\sim$5 Mbps to get back the binocular videos; here, we also regarded the left views as the monocular videos.
%
	\item[(iv)] \emph{Control-group B}. From the ground truths, we encoded and decoded the left view of them by the H.264 codec at low bit-rate ($\sim$1 Mbps) to slightly inject some visual artifacts, such as noise and blurriness in the video frames; then, we randomly removed some frames in the videos to create slight temporal discontinuity as the monocular videos.
	Further, we shifted the generated monocular videos some pixels to \new{the left (as shown in Figure 8 in the supplementary material), to} act as the right view videos, and regarded the monocular videos together with the shifted videos as the binocular videos.
%
\end{itemize}
Control-group A is set for simulating the conventional video quality of monocular and binocular platforms, while control-group B is set for checking whether the users carefully participate in the study.
Note also that the videos in control-group B are not obviously bad; see Supplemental material Section 5 for examples.
%

%
Altogether, we prepared eight sets of videos (ground truths, our framework (ours), control-group A, and control-group B) for monocular videos, and another eight sets for binocular videos.
Concerning the participants, we recruited 25 participants: 10 females and 15 males, all with normal vision, and aged 24 to 28.


\vspace{-3pt}
\paragraph{Procedure}
Our user study has three sessions.
The first one is a tutorial session.
When a participant came to our lab, we first showed to him/her some normal videos and some control videos (like those in control-group B) for both monocular and binocular.
This was to ensure that the participants knew the meaning of visual artifacts, temporal smoothness, and depth perception, and could perceive depth, when watching the binocular videos.
Here, we employed the \textit{Bino Player}\footnote{Bino Player:~\url{https://bino3d.org/}} to show binocular videos on a 27'' polarized 3D display\new{, with the  resolution of $1920\times1080$ and the peak luminance of  $250cd/m^2$}.

The second session focused on the eight sets of monocular videos.
Here, we showed the videos to each participant set by set, in which the four video types in each set (i.e., ground truths, ours, control-group A, and control-group B) were shown in random order.
To avoid bias, we use different random orders for different sets.
After seeing all the four videos in a set, the participant might go back and forth in the playlist to carefully examine the four videos again before they gave a rating in the scale of one (poor quality) to five (excellent quality) per video for questions Q1 to Q3 listed in Figure~\ref{fig:userStudy}.

The third session followed a similar procedure as in the second session, but we showed the other eight sets of binocular videos on the 3D display and asked for ratings on questions Q4 to Q7 (Figure~\ref{fig:userStudy}).
%
%
After the three sessions, we obtained a total of
``25 participants
$\times$
8 video samples
$\times$
4 video types''
per question.


\begin{figure*}[!t] 
	\centering
	\includegraphics[width=0.99\linewidth]{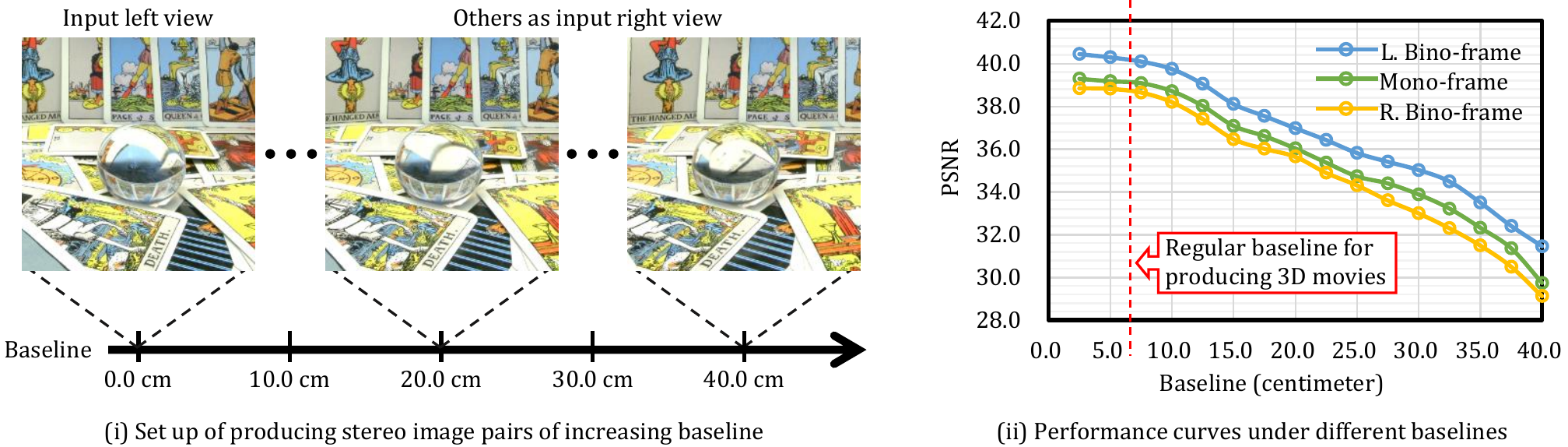}
	\vspace*{-2mm}
	\caption{
	\new{
	Stress test on our method using stereo image pairs of increasing ``baseline'' distances.
	(i) illustrates the set up, in which we treated the left-most image as the input left view and other images as the input right view.
	(ii) plots the quality of our results for stereo image pairs of increasing baseline.
	}	
	}
	\label{fig:breakpointn}
\end{figure*}

\vspace{-3pt}
\paragraph{Analysis of the results}
Figure~\ref{fig:userStudy} (left) summarizes the mean participant ratings for questions Q1 to Q3 on the four types of monocular videos.
Comparing the ratings on control-group B vs. others, we can clearly see that the participants could distinguish the control-group B videos from others.
%
Comparing the ratings on ground truths, ours, and control-group A,
we can see that their rating distributions are very similar for all the three questions Q1 to Q3.
To statistically compare them, we performed an equivalence test using the two-one-sided t-test (TOST)~\cite{schuirmann1987comparison}, because t-test can only help to examine significant difference between two groups of data, while equivalence test can tell whether the two groups are equivalent under a given bound, which is specified to denote the smallest effect size of interest.
Here, the upper and lower equivalence bounds in TOST are set to be $0.5$ and $-0.5$, respectively, since the participant ratings are all integers and 0.5 is half the interval size.
The result of the TOST shows that the ratings on ours and control-group A for questions Q1-Q3 are equivalent with confidence values $99.2\%$, $99.5\%$, and $99.7\%$, respectively.
On the other hand, the confidence values are $99.4\%$, $98.2\%$, and $99.4\%$, respectively, for the ratings on ours and ground truths for Q1-Q3.
Hence, the results suggest that the ratings on ours and ground truths for Q1-Q3 are equivalent, meaning that there are no obvious perceptual differences between the original (left view) videos and our mononized videos.
%

Next, we look at Figure~\ref{fig:userStudy} (right) for the binocular videos.
Again, the control-group B videos can be recognized by the participants, and the rating distributions for ground truths, ours, and control-group A are similar for questions Q4 to Q7.
Using TOST with the same setting as before, we found that the ratings on ours and control-group A for Q4 to Q7 are equivalent with confidence values $99.2\%$, $98.7\%$, $99.9\%$, and $99.6\%$, respectively, whereas the ratings on ours and ground truths for Q4 to Q7 are equivalent with confidence values $97.9\%$, $98.6\%$, $99.1\%$, and $98.5\%$, respectively.
Hence, there are no obvious perceptual differences, both between ours and control-group A and between ours (our restored binocular videos) and ground truths.
The detailed questionnaire and example frames can be found in Section 5 of the supplemental material.

\subsection{Discussion}
\label{subsec:limitation}

\new{
\paragraph{How does the performance vary with inter-camera distance?}
The interpupillary distance, or the distance between the centers of the pupils of the eyes, varies from 5.1 to 7.7 cm for adults, whereas the average is 6.2 cm for females and 6.4 cm for males\footnote{\url{https://en.wikipedia.org/wiki/Pupillary_distance}}.
Hence, for general 3D movies that aim to reproduce natural human vision, the distance between camera centers are often set to a ``normal'' baseline of 5 to 8 cm\footnote{\url{https://en.wikipedia.org/wiki/Stereo_photography_techniques}}.
%
%
%
%
Obviously, the task of mononizing binocular videos would become more challenging when the baseline increases and when some objects are close to the cameras.
It is because doing so will increase the disparity between the left and right views, thus making it harder to encode the two views into one.
We performed a stress test on our method by
setting up the scene shown in Figure~\ref{fig:breakpointn}(i).
It is a very challenging scene, since the crystal ball shown introduces strong non-Lambertian reflections and discontinuity, while being close to the viewpoints.
Then, we tested the performance of our method on a sequence of stereo image pairs captured at viewpoints of increasing baselines, in which we treated the left-most image (at 0.0 cm) as the input left view and other images of increasing baselines from the left as the input right views.
%
%

%
%
Figure~\ref{fig:breakpointn} (ii) plots the quality of the mononized and restored binocular frames produced by our method on stereo image pairs of increasing baselines.
We can see from the plots that the quality of our results stays very high ($>$ 38 dB) and varies only slightly for baseline less than 10 cm.
When the baseline increases from 10 to 35 cm, the quality smoothly decreases but is still well above 30 dB.
Further, when the baseline reaches 40 cm, the quality of the mononized and restored right views drops below 30 dB, for which the audiences may be able to aware of the artifacts.
%
%
}

\paragraph{How is the stereo information encoded?}
\begin{figure}[!t] 
	\centering
	\includegraphics[width=0.99\linewidth]{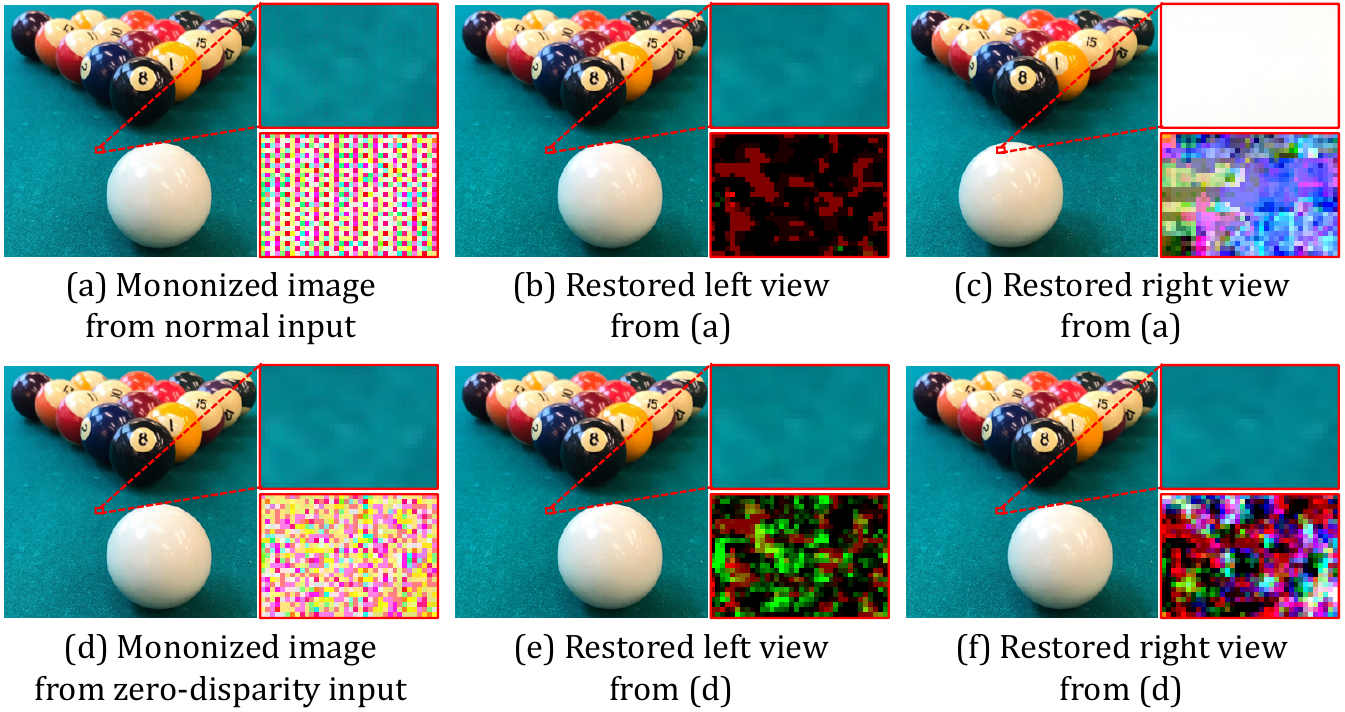}
	\vspace*{-2mm}
	\caption{
		The first row shows the mononized view (a) produced from a normal binocular input and left \& right views (b \& c) restored from it.
		The second row shows the mononized view (d) produced from a zero-disparity input and left \& right views (e \& f) restored from it.
		The right-hand sides of (a)-(f) present the blown-up regions (top) and difference images from the ground truths (bottom), which have been scaled-up ``100'' times for viewing.
	}
	\label{fig:stereo_pattern}
	\vspace*{-4mm}
\end{figure}
In our investigation, we first extensively zoom into ($\sim$30$\times$) a small region in a mononized view that is occluded in the corresponding right view (Figures~\ref{fig:stereo_pattern} (a) \& (c)).
We choose such a region, since the left and right views in the region are substantially different, so more stereo information has to be encoded in the region inside the mononized view.
However, we cannot observe any abnormal pattern in the blown-up view of this small region (top-right image in Figure~\ref{fig:stereo_pattern} (a)).
Hence, we further compute the difference image in the region between the mononized view and input left view, and scale up the difference ``100'' times for better visualization.
The bottom-right image in Figure~\ref{fig:stereo_pattern} (a) shows the result, in which a regular dot pattern is revealed.

Intuitively, we suppose the stereo information is carried by the dot pattern.
After the restoration, the dots no longer exist in the restored left and right views (b \& c), since our decoding network has consumed them in the restoration process.
To further verify our supposition, we feed a zero-disparity image pair (i.e., left view = right view) to our {\em encoding} network, to produce a mononized view (d), as well as restoring from it a pair of left and right views (e \& f).
The dot patterns are absent in these results, thus revealing that the regular dots indeed encode the stereo information.

\begin{figure}[!t] 
	\centering
	\includegraphics[width=0.99\linewidth]{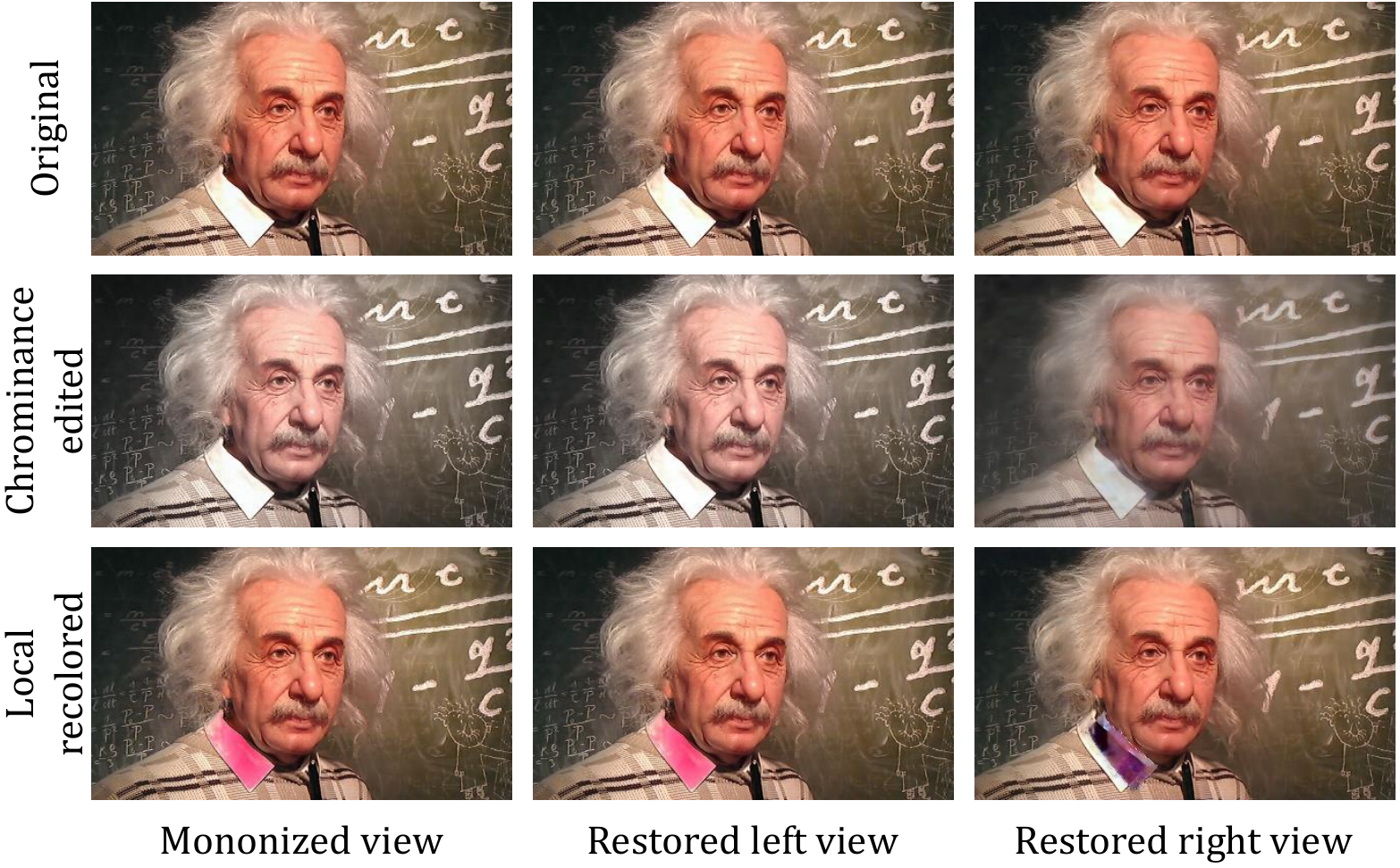}
	\vspace{-2mm}
	\caption{Effects of manipulating the mononized view. Left column shows the original and manipulated mononized views, while the middle and right columns show the corresponding restored left and right views, respectively.
	}
	\vspace*{-4mm}
	\label{fig:limitation}
\end{figure}

\begin{figure*}[htbp] 
	\centering
	\includegraphics[width=0.95\linewidth]{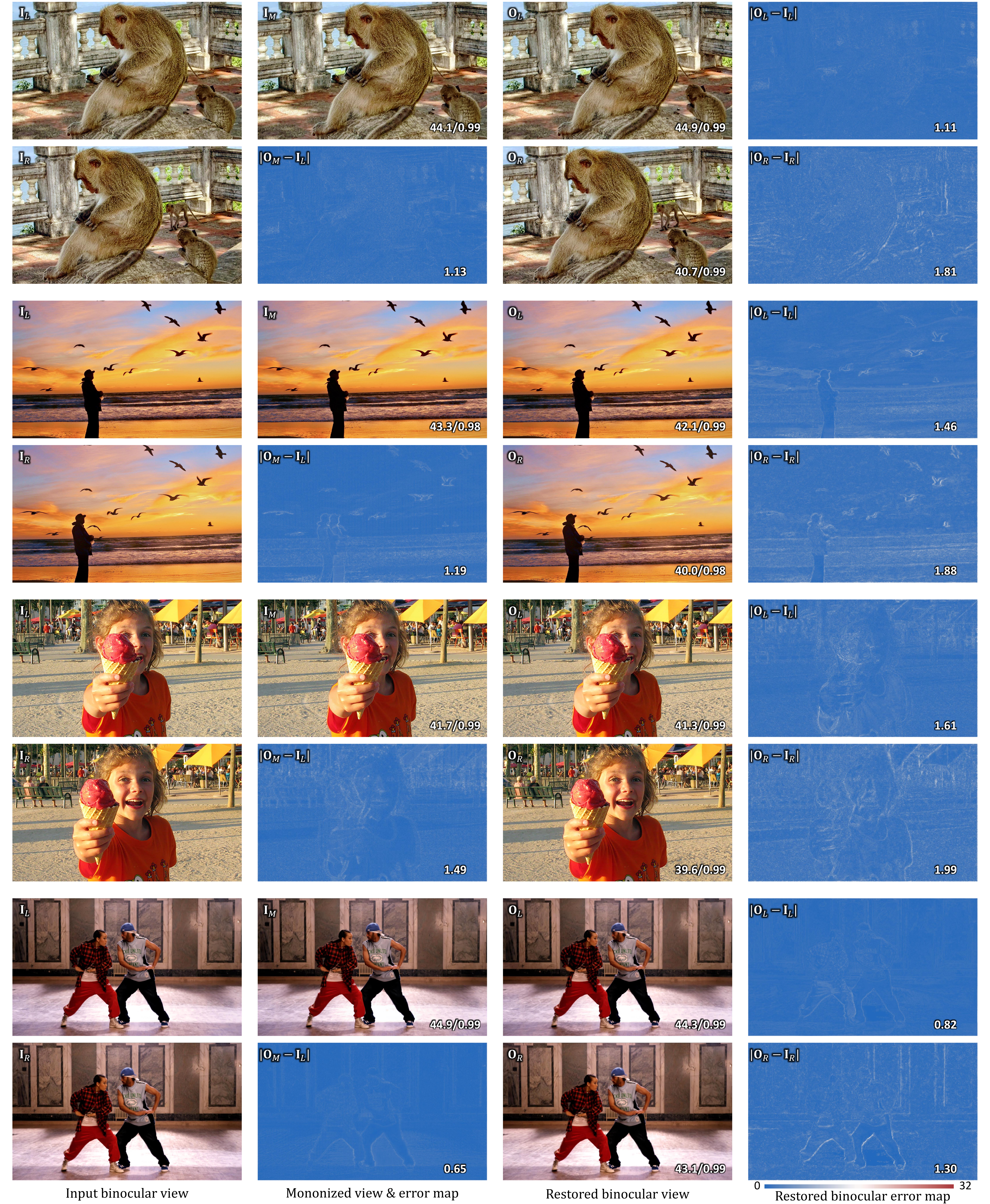}
	\vspace{-4mm}
	\caption{Four sets of example results (every pair of rows) produced by our method.
		In each result, we show the input left \& right views (1st column), generated mononized view and its difference map from the input left view (2nd column), restored binocular views (3rd column), and their difference maps from the inputs (4th column).
		In each result, the numbers show PSNR and SSIM, while in each error map, the number shows the mean absolute difference (scale of [0,255]).}
	\label{fig:showcase}
\end{figure*}

\paragraph{Limitation}
Manipulations on the mononized view can hurt the restorability.
For example, if we edit the chrominance in the mononized image, the manipulation effect could be somehow transferred to the restored left view but the restored right view could suffer from blurriness (second row in Figure~\ref{fig:limitation}).
Similarly, locally re-coloring objects in the mononized image could lead to improper color changes in the corresponding image region in the right view (third row in Figure~\ref{fig:limitation}).
It is because the manipulations ruin the stereo information visually-encoded in the mononized view, thus interfering the restoration of the binocular views by the decoding network.

\section{Conclusion}
\label{sec:conclusion}

We presented an innovative idea of {\em mono-nizing\/} binocular videos and a framework to effectively realize it by implicitly encoding the stereo information in a visual but nearly-imperceptible form inside the mononized videos.
Our mononized videos allow us not only to impartially distribute and show them as ordinary monocular videos on existing video platforms but also to decode them back to binocular videos for stereo viewing, when a 3D display is available.
Our technical contributions include an encoding-and-decoding framework 
with the pyramidal deformable fusion module to exploit long-range correspondences between the left and right views,
a quantization layer to suppress the 
restoring 
artifacts, and 
the compression noise simulation module to resist the compression noise introduced by modern video codecs.
Our framework is self-supervised.
We formulate the objective with a monocular term, an invertibility term, and a temporal term, which are defined on the input binocular video to guide the network to produce the mononized and binocular videos. 
%
%
Extensive experiments were performed to show the quality of our results and backward compatibility of our method.
%
Particularly, our mononized videos and restored binocular videos look no different from the original ones, and our mononized videos are compatible with various common video codecs, as demonstrated in the user study and experiments.

%

%
In the future, we would like to further extend our framework to mononize general multi-view videos acquired on the various recently-launched multi-camera phones.
Also, we are interested in exploring editable mononized images/videos, such that global and local image manipulations applied on the mononized view can be transferred to all of the restored multi-views.
Having editable mononized images/videos could enable the manipulation of multi-view scenes in a coherent and efficient manner.

\begin{acks}
\new{This project is supported by the Research Grants Council of the Hong Kong Special Administrative Region, under RGC General Research Fund (Project No. CUHK14201017 and CUHK14201918).
The original binocular photo in Figure 1 is from \emph{3D shoot} (Flickr); it is licensed under CC BY-NC-SA 2.0.}
\end{acks}

\bibliographystyle{ACM-Reference-Format}
\bibliography{InvertBino}

\end{document}